%% file: lens100d.tex
\documentclass[apj]{emulateapj}

\usepackage{amsmath}
\usepackage{natbib}
\usepackage{graphicx}
\usepackage{epsf}
\usepackage{color}
\usepackage{threeparttable}
\usepackage{comment}
\usepackage{epsfig}
\usepackage{xspace}
\usepackage{subfigure}
\usepackage{ulem} 
\usepackage[usenames,dvipsnames,svgnames]{xcolor}
\DeclareGraphicsExtensions{.jpg,.pdf,.png,.eps,.ps}

\newcommand{\mvSysTcal}{\ensuremath{0.052}}
\newcommand{\mvSysPcal}{\ensuremath{0.030}}
\newcommand{\mvSysXtalk}{\ensuremath{0.05}}
\newcommand{\mvSysAmp}{\ensuremath{0.029}}

\newcommand{\ppSysTcal}{\ensuremath{0.052}}
\newcommand{\ppSysPcal}{\ensuremath{0.067}}
\newcommand{\ppSysXtalk}{\ensuremath{0.06}}
\newcommand{\ppSysAmp}{\ensuremath{0.039}}

\newcommand{\mvAmp}{\ensuremath{0.92}}
\newcommand{\mvAmpStat}{\ensuremath{0.14}}
\newcommand{\mvAmpSys}{\ensuremath{0.08}}
\newcommand{\mvAmpUnlStat}{\ensuremath{0.065}}

\newcommand{\ppAmp}{\ensuremath{0.92}}
\newcommand{\ppAmpStat}{\ensuremath{0.24}}
\newcommand{\ppAmpSys}{\ensuremath{0.11}}

\newcommand{\mvSig}{\ensuremath{6.5 \sigma}}        
\newcommand{\mvSigUnl}{\ensuremath{14 \sigma}}    
\newcommand{\mvPercent}{\ensuremath{15 \%}}      
\newcommand{\ppSigUnl}{\ensuremath{5.9 \sigma}}     
\newcommand{\ppPercent}{\ensuremath{26 \%}}      

\newcommand{\mvDampTot}{\ensuremath{0.16}}     
\newcommand{\ppDampTot}{\ensuremath{0.26}}     

\newcommand{\mvTotPercent}{\ensuremath{18 \%}}      
\newcommand{\ppTotPercent}{\ensuremath{29 \%}}      

\newcommand{\mvPTE}{\ensuremath{0.49}}
\newcommand{\ppPTE}{\ensuremath{0.10}}

\newcommand{\mvLCDMSig}{\ensuremath{0.47 \sigma}}
\newcommand{\ppLCDMSig}{\ensuremath{0.32 \sigma}}

\newcommand{\nbin}{\ensuremath{10}}

 \newcommand{\nhat}{\ensuremath{\mathbf{\hat{n}}}}

 \newcommand{\sz}{Sunyaev-Zel'dovich}

 \newcommand{\LCDM}{\mbox{$\Lambda$CDM}\xspace}
 \newcommand{\muksq}{\ensuremath{\mu{\rm K}^2}}

 \newcommand{\wmap}{\textit{WMAP}\xspace}
 
 \newcommand{\sptpol}{SPTpol}
 \newcommand{\planck}{\textit{Planck}}
 \newcommand{\polarbear}{POLARBEAR}
 \newcommand{\bicep}{BICEP2}
 \newcommand{\keck}{KECK Array}
 \def\muKarcmin{$\mu{\mbox{K}}$-arcmin}

 \def\mb{\mathbf}     
 \def\mbL{\mathbf{L}} 
 \def\mbell{\pmb{\ell}} 

 \def\clpp{\ensuremath{C^{\phi \phi}_L}}
 
 \def\clppHatUVXYL{\ensuremath{C^{\hat{\phi}^{UV} \hat{\phi}^{XY}}_{\mbL}}}
 \def\cmbLppHat{\ensuremath{C^{\hat{\phi} \hat{\phi}}_{\mbL}}}
 \def\hatcmbLpp{\ensuremath{\hat{C}^{\phi \phi}_{\mbL}}}
 
 \def\hatclppUVXYL{\ensuremath{\hat{C}^{\phi^{UV} \phi^{XY}}_{\mbL}}}

 \def\dclNzero{\ensuremath{\left. \Delta C_{\mbL}^{\phi \phi}\right|_{\rm N0}}}
 \def\dclRDNzero{\ensuremath{\left. \Delta C_{\mbL}^{\phi \phi}\right|_{\rm RDN0}}}
 \def\dclNone{\ensuremath{\left. \Delta C_{\mbL}^{\phi \phi}\right|_{\rm N1}}}
 \def\dclMC{\ensuremath{\left. \Delta C_{\mbL}^{\phi \phi}\right|_{\rm MC}}}
 
 \def\mv{MV}
 \def\pp{POL}

\def\be{\begin{equation}}
\def\ee{\end{equation}}
\def\ba{\begin{eqnarray}}
\def\ea{\end{eqnarray}}

\begin{document}
\title{A Measurement of the Cosmic Microwave Background Gravitational Lensing Potential from 100 Square Degrees of SPTpol Data}

\input lens100d_authorlist_v1.tex

\begin{abstract}
We present a measurement of the cosmic microwave background (CMB) gravitational lensing potential using data from the first two seasons of observations with \sptpol, 
the polarization-sensitive receiver currently installed on the South Pole Telescope (SPT).
The observations used in this work cover 100 deg$^2$ of sky with arcminute resolution at $150\,$GHz.
Using a quadratic estimator, we make maps of the CMB lensing potential from combinations of CMB temperature and polarization maps.
We combine these lensing potential maps to form a minimum-variance (\mv) map.
The lensing potential is measured with a signal-to-noise ratio of greater than one for angular multipoles between $100< L <250$.
This is the highest signal-to-noise mass map made from the CMB to date and will be powerful in cross-correlation with other tracers of large-scale structure.
We calculate the power spectrum of the lensing potential for each estimator, and we report the value of the \mv{} power spectrum between $100< L <2000$ as our primary result.
We constrain the ratio of the spectrum to a fiducial \LCDM{} model to be 
$A_{\rm \mv}=\mvAmp \pm \mvAmpStat {\rm\, (Stat.)} \pm \mvAmpSys {\rm\, (Sys.)}$.
Restricting ourselves to polarized data only, we find 
$A_{\rm \pp}=\ppAmp \pm \ppAmpStat {\rm\, (Stat.)} \pm \ppAmpSys {\rm\, (Sys.)}$.
This measurement rejects the hypothesis of no lensing at \ppSigUnl{} using polarization data alone,
and at \mvSigUnl{} using both temperature and polarization data.
\end{abstract}

\keywords{cosmic background radiation -- gravitational lensing -- large-scale-structure}

\maketitle


\section{Introduction}
\setcounter{footnote}{0}

Gravitational lensing of the cosmic microwave background (CMB) has become a powerful observational tool for probing the geometry of the universe and the growth of large-scale structure.
As light travels through the universe, the paths of photons are bent by gravitational interactions with matter.
These deflections encode information about the growth of large-scale structure at late times, when massive neutrinos suppress structure growth and dark energy drives the accelerated expansion of the universe.
As a result, lensing measurements provide a valuable method for constraining neutrino masses \citep{lesgourgues06b} and dark energy \citep{calabrese09}.
The CMB provides a unique source for lensing measurements:
the CMB is well characterized,
originated at a well known redshift,
and has traversed nearly the entire observable universe, in principle providing information about all of the structure between Earth and the last scattering surface of the CMB.
The lensing signal is sourced from a broad range of redshifts, typically $0.1<z<5$ \citep{lewis06}.

We can measure gravitational lensing by looking for distortions in the characteristic patterns of the primordial CMB temperature and polarization (\citealt{blanchard87}; for a review, see \citealt{lewis06}).
The CMB is partially polarized, and can be decomposed into even-parity (E-mode) and odd-parity (B-mode) components \citep[e.g.,][]{hu97d}.
Lensing stretches and contracts observed CMB anisotropies.
These distortions smooth the CMB temperature \citep{seljak96b} and E-mode power spectra, 
produce lensed B-mode power on scales where there is very little primordial power \citep{zaldarriaga98},
and correlate modes that were originally independent \citep{cole89b}.
Lensing in the CMB was first detected using lensing-galaxy correlations \citep{smith07a, hirata08}.
The smoothing of the CMB power spectrum has subsequently been well measured in temperature \citep{story13, das14, planck13-16}
and should soon be detected in the E-mode power spectrum.

Lensing-induced B-modes were first detected in cross-correlation using data from \sptpol{},  the polarization-sensitive receiver currently installed on the South Pole Telescope \citep[SPT,][]{carlstrom11};
\sptpol{} data was cross-correlated with the cosmic infrared background, resulting in a $7.7\sigma$ detection of B-modes \citep[hereafter H13]{hanson2013}.
Using similar cross-correlation analyses, 
\polarbear{} found $2.3\sigma$ evidence for lensed B-modes (\citeauthor{polarbear2014c}),
and the Atacama Cosmology Telescope Polarimeter (ACTPol) collaboration measured lensed B-modes with a $3.2\sigma$ significance \citep{vanengelen14b}.
Recently, \polarbear{} made a measurement of the B-mode auto-spectrum which disfavored the no-lensing hypothesis at $2.0\sigma$ (\citeauthor{polarbear2014b}),
and \bicep{} measured a B-mode auto-spectrum which contains a lensing component measured at $5.5\sigma$ \citep{bicep2a}.

The correlation between initially independent CMB modes introduced by lensing can be measured using a quadratic estimator technique \citep{seljak99, hu01b, hu02a}.
With this technique, the amplitude of the lensing potential power spectrum has been measured to high precision using CMB temperature data:
$22\%$ using data from $\sim 600$ deg$^2$ of sky observed with the Atacama Cosmology Telescope (ACT) \citep{das14},
$19\%$ using data from $\sim 590$ deg$^2$ of the 2500 deg$^2$ SPT-SZ survey observed with the SPT \citep{vanengelen12}, and 
$4\%$ using data from $\sim 28000$ deg$^2$ of sky ($70\%$ of the full sky) observed with \planck{} \citep{planck13-17}\footnote{
The fractional precision is reported relative to the mean of each measurement.}.
Data from the full 2500 deg$^2$ SPT-SZ survey is expected to yield a measurement of the lensing amplitude with a significance similar to that of the current \planck{} result.
\polarbear{} has detected gravitational lensing with a quadratic-estimator technique using only estimators that include $B$-modes, 
 rejecting the no-lensing hypothesis at $4.2\sigma$ over 30 deg$^2$ (\citeauthor{polarbear2014a});
combining this with their B-mode auto-spectrum measurement, \polarbear{} rejected the no-lensing hypothesis at $4.7\sigma$ (\citeauthor{polarbear2014b}).

In this paper, we construct a map of the gravitational lensing potential and its power spectrum from temperature and polarization data collected with \sptpol{} within a 100 deg$^2$ patch of the southern sky.
We combine the temperature and polarization measurements into a minimum-variance (\mv) estimate of the lensing potential.

Throughout this paper, we use a fiducial \LCDM{} cosmological model that provides the best fit to the combination of \planck{} power spectrum, \planck{} lensing spectrum, \wmap-polarization, SPT, and ACT data \citep{planck13-16},
with the following parameters:
baryon density $\Omega_b h^2=0.0222$,
cold dark matter density $\Omega_c h^2=0.1185$, 
Hubble parameter $H_0=100h$km s$^{-1}$ Mpc$^{-1}$ with $h=0.6794$,
power spectrum of primordial curvature perturbations with an amplitude (at $k=0.05$ Mpc$^{-1}$) $A_s=2.21\times 10^{-9}$
and spectral index $n_s=0.9624$,
optical depth to reionization $\tau=0.0943$,
and a neutrino energy density corresponding to a sum over the neutrino mass eigenstates of $0.06$ eV.
Hereafter, this model will be referred to as the ``\textsc{Planck+Lens+WP+highL}'' model.
This fiducial model serves as the input to our simulations 
and the reference for calculating the amplitude of the lensing potential power spectrum.

This paper is organized as follows:
We describe the data used in this paper in Section \ref{sec:data}.
We review the theoretical description of gravitational lensing and describe our implementation of the quadratic estimator in Section \ref{sec:theory}.
We describe the simulations in Section \ref{sec:sims}.
We detail our uncertainty budget in Section \ref{sec:errors}.
We present our results in Section \ref{sec:results}.
We discuss potential sources of systematic errors in Section \ref{sec:systematics}.
We discuss the implications of this measurement in Section \ref{sec:discussion}.

\section{Sptpol Observations and Data-Processing}
\label{sec:data}

This paper uses maps of the CMB temperature and polarization 
made with data from \sptpol{} \citep{austermann12}.
\sptpol{} is a polarization-sensitive receiver that was installed in early 2012 on the 10-meter South Pole Telescope \citep{carlstrom11}. 
We use data obtained from observations during 2012 and a few months of 2013.
The 2012 data have already been used to make the first detection of lensing B modes (H13) 
and to make the most precise measurement of the high $\ell$ E-mode power spectrum and temperature-E-mode cross-spectrum to date \citep[hereafter C14]{crites14}.
A more thorough discussion of the \sptpol{} instrument and data set is given in C14; here we summarize the most important properties of the \sptpol{} data and resulting maps.

The \sptpol{} receiver contains 1536 polarization-sensitive transition edge sensor bolometers, with 1176 detectors at 
$150\,{\rm GHz}$ and $360$ detectors at $95\, {\rm GHz}$.
In this analysis we use only the $150\,{\rm GHz}$ data. 
During 2012 and the first part of 2013, \sptpol{} was used to observe the $~100\ {\rm deg}^2$ ``\sptpol{} deep field,'' 
a roughly-square patch of sky spanning right ascensions of 23h to 24h and declinations of $-50$ to $-60$ degrees.
The data consists of 
5182 
individual field observations, each lasting approximately 30 minutes. 
The observations were performed using a lead-trail observing strategy; 
the field is split into two halves (``lead'' and ``trail'') in right ascension, which are scanned separately such that each half-field is observed over the same range in azimuth. 
The half-fields are observed by scanning the telescope back and forth in azimuth, followed by a small step in elevation;
this process is repeated until the entire half-field is covered.
We refer to a single sweep of the telescope from one side of the field to the other as a ``scan,''
and to a set of scans that cover the entire half-field as an ``observation.''
The lead-trail strategy allows the subtraction of ground pickup by differencing pairs of lead and trail observations, 
however, in this work we simply coadd lead and trail observations into single observation maps.

The raw field observations consist of time-ordered data (TOD) for each \sptpol{} bolometer. 
The individual bolometer TOD are calibrated relative to one another using a combination of regular calibration observations of 
an internal chopped blackbody source and of the galactic HII region RCW38. 
The absolute calibration ($T_{\rm cal}$) is tied to the CMB maps produced by \planck. 
The absolute temperature calibration uncertainty is estimated to be
$\delta_{\rm Tcal} = 1.3\%$.
See C14 for a detailed description of this calibration process.

The TOD are filtered before being accumulated into maps. 
To suppress atmospheric fluctuations, a fourth-order polynomial is subtracted from the TOD of each detector for every scan. 
A low-pass filter is also applied to prevent aliasing at the pixelization scale,  
with a cutoff corresponding roughly to a maximum angular scale of 
$\ell \sim 4000$ ($\sim 3$ Hz at the telescope scan-speed).

Maps are made by calculating the Stokes parameters $I$, $Q$, and $U$ in each map pixel.
We refer to the Stokes parameter $I$ as $T$ (temperature), and use temperature units for our $T$, $Q$, and $U$ maps.
The TOD are accumulated into maps using the pointing, polarization angle, and efficiency of each detector,
as well as a weight calculated from the noise power of each detector between 1 and 3 Hz ($1300 \lesssim \ell \lesssim 3900$ at the telescope scan-speed).
The maps are pixelized in an oblique Lambert azimuthal equal-area projection, 
with square $2' \times 2'$ pixels.

The $Q$ and $U$ maps are multiplied by an additional calibration factor $P_{\rm cal}$ to correct for errors in the measured detector polarization calibration and efficiency.
We use a value of $P_{\rm cal}=1.048$ from Table 3 in C14.
This value is calculated using a Markov Chain Monte Carlo (MCMC)\footnote{
We calculate the MCMC chain using the \textsc{CosmoMC} package \citep{lewis02b}.
}
by fitting the \sptpol{} $EE$ and $TE$ bandpowers jointly with \planck{} $TT$ bandpowers \citep{planck13-16} and \textit{WMAP}9 polarization data \citep{hinshaw13} 
in a \LCDM{} model with varying nuisance parameters; see C14 for details.
The value of $P_{\rm cal}$ for the best-fit model from this MCMC-chain is taken as our estimate of the polarization calibration parameter.
The $P_{\rm cal}$ calibration has an uncertainty of $\delta_{\rm Pcal}=1.7\%$, which is calculated from the marginalized posterior of $P_{\rm cal}$.
We treat $\delta_{\rm Pcal}$ as uncorrelated with other systematic uncertainties in Section \ref{sec:errors};
this over-estimates the uncertainty since $P_{\rm cal}$ is partially correlated with other parameters in the fit (e.g., $T_{\rm cal}$).

The final \sptpol{} deep field maps have nearly homogeneous coverage, with inverse-noise-weighted pixel hit counts varying by less than $\sim35\%$ over the 100 deg$^2$ field (corresponding to the difference in the cosine of the declination angle between the top and bottom of the field), 
except for a $\sim 2\times$ deeper strip at the center caused by overlap of the lead and trail fields.
The maps have an effective noise level, estimated between $2000 < \ell < 3000$,
of 
$\sim 11$ \muKarcmin{} in temperature and $\sim 9$ \muKarcmin{} in $Q$ and $U$
(atmospheric noise causes a higher noise level in $T$ than in $Q$ or $U$).
We use data-quality checks to cut data at three stages: individual bolometers, scans, and observations.
These cuts are described in detail in C14.
All data that pass these cuts are coadded into final $T$, $Q$, and $U$ maps.

Systematic effects such as detector gain errors can cause temperature power to leak into the $Q$ and $U$ polarization maps.
We correct for the leakage by subtracting the appropriately scaled temperature map from each polarization map as follows:
$Q = Q_{raw} - \hat{\epsilon}^Q T$, and an analogous expression for $U$.
We find 
$\hat{\epsilon}^Q = +0.0050$ and 
$\hat{\epsilon}^U = -0.0083$.
See C14 for more detail.

Signal power in the final maps is suppressed by two processes: data filtering and the 
telescope angular response function, or beam.
We model the signal power that is removed by filtering as a 2D Fourier-space\footnote{
Throughout this paper, we adopt the flat-sky approximation, 
where wavenumbers are represented by the discretized two-dimensional  vector $\mbell$,
the magnitude of $\mbell$ is equivalent to multipole number $\ell$, and spherical harmonic transforms are replaced by Fourier transforms.}
transfer function $F^{\rm filt}_{\mbell}$.
We obtain this transfer function using simulations of the filtering process.
The SPT beam suppresses power on arcminute scales.
We approximate this beam as radially symmetric, with a 2D Fourier-space transfer function $F^{\rm beam}_{\mbell}$ measured using observations of Mars as well as bright point sources in the deep field.
The beam was measured independently in both 2012 and 2013;
we use a composite beam calculated as the inverse-noise-variance weighted average of the beams from each year.
%
Taken together, we model the total transfer function as 
\be
\label{eq:tf}
F^{\rm tot}_{\mbell} = F^{\rm pix}_{\mbell}\, F^{\rm filt}_{\mbell}\, F^{\rm beam}_{\mbell}
\ee
where $F^{\rm pix}_{\mbell}$ is the 2D Fourier transform of a square $2\arcmin$ pixel.

The \sptpol{} deep field contains a number of bright point sources, which we mask.
We identify all sources detected at $>5\sigma$ at $150\,{\rm GHz}$ from \cite{mocanu13},
which corresponds to a flux-cut of approximately $6\,$mJy;
all pixels within $5\arcmin$ of each source are masked in both the TOD filtering and the final maps.
We extend this mask to $10\arcmin$ for all very bright sources detected at $>75\sigma$. 
We also use a $10\arcmin$ radius to mask all galaxy clusters detected in this field by \cite{vanderlinde10} using the \sz\ effect.
This masking removes approximately 
$120$ sources, cutting
5 deg$^2$ of the field.
%
%
%
We additionally multiply the maps by a sky mask that down-weights the noisy edges of the maps.

The processing described in this section results in a set of three coadded, masked maps: $T(\hat{\mb{n}})$, $Q(\hat{\mb{n}})$, and $U(\hat{\mb{n}})$.
These maps are filtered with the C-inverse filtering process (see Section \ref{sec:cinv}),
which transforms the set of maps into filtered Fourier-space maps $\bar{T}_{\mbell}$, $\bar{E}_{\mbell}$, and $\bar{B}_{\mbell}$.

We apply two cuts in Fourier-space.
The data maps have noise that arises from two contributions: 
a roughly isotropic component from atmosphere at low $|\mbell|$ 
and an anisotropic component from low-frequency noise that is uncorrelated between detectors at low $|\ell_x|$. 
We remove these noisy regions of Fourier space, keeping only modes with $|\ell_x| > 450$.
Second, we use only modes with $|\mbell| < 3000$ when estimating the lensing potential.
We verify that our results are not sensitive to the exact choices of these cuts; see Section \ref{sec:systematics} for details.

Thus the data input to the estimator of the lensing potential (i.e., Equation \ref{eq:phi_bar}) is a set of three filtered maps: $\bar{T}_{\mbell}$, $\bar{E}_{\mbell}$, and $\bar{B}_{\mbell}$.

\section{CMB Lensing Analysis}
\label{sec:theory}

As CMB photons travel through the universe, their paths are deflected by gravitational interactions with intervening large-scale structure.
These deflections can be described as a remapping of the CMB temperature and polarization fields \citep{lewis06}:
\ba
\label{eq:lens_grad}
T(\nhat) &=& \tilde{T}(\nhat + \nabla\phi(\nhat)) \\
\left[Q \pm iU \right](\nhat) &=& [\tilde{Q} \pm i\tilde{U}](\nhat + \nabla\phi(\nhat)) \,,
\ea
where $\nhat$ is a unit-vector denoting a particular line-of-sight direction on the sky.
Throughout the paper, tildes denote \textit{unlensed} quantities.
The projected lensing potential $\phi(\nhat)$ is an integral over the 3-dimensional potential $\Psi(\chi \nhat; \eta_0 - \chi )$:
\be
\phi(\nhat) = -2 \int_0^{\chi_{\rm CMB}} d\chi
\frac{ f_{K}\left(\chi_{\rm CMB} - \chi \right)}{f_{K}\left(\chi_{\rm CMB}\right) f_{K}\left(\chi\right)}
\Psi(\chi \nhat; \eta_0 - \chi ).
\label{eq:phin}
\ee
where $\chi$ is the comoving distance along the line of sight, 
$\chi_{\rm CMB}$ is the comoving distance to the surface of last scattering, 
the quantity $\eta_0 - \chi$ is the conformal time at which a CMB photon would have been at position $\chi \nhat$,
and $f_{K}(\chi)$ is the comoving angular distance with $f_{K}(\chi)=\chi$ in a flat universe.
The deflection field of CMB photons is given by $\vec{d}(\nhat) = \nabla\phi(\nhat)$.
The apparent local expansion or contraction of the CMB is given by the divergence of $\nabla\phi$;
the convergence field is defined as $\kappa(\nhat) \equiv -\frac{1}{2} \nabla^2 \phi(\nhat)$.
In this paper, final maps and power spectra will be presented in terms of $\kappa$.

The unlensed CMB sky is well described as a statistically-isotropic, Gaussian random field; in this case, all statistical information about the field is contained in the power spectrum 
\[C_{\ell}^{XY} = \delta(\mbell - \mbell')\langle X_{\pmb{\ell\phantom{'}}} Y^{*}_{\mbell'} \rangle\,,\] 
where $X, Y \in [T,E,B]$ are harmonic-space CMB maps.
Gravitational lensing introduces statistical anisotropy in the CMB temperature and polarization fields, correlating multipole moments across a range defined by the lensing deflection field $\vec{d}(\nhat)$ \citep{hu02a}.
Averaging over realizations of the lensed CMB fields for a fixed lensing potential, 
we can expand the covariance of CMB fields in a Taylor series as a function of $\phi$ \citep{hu02a}:
\be
\label{eq:lens_expansion}
\langle X_{\pmb{\ell\phantom{'}}} Y^{*}_{\pmb{\ell'}} \rangle_{\rm CMB} = \delta(\mbell-\mbell') C_{l}^{XY} + W^{\phi^{XY}}_{\mbell, \pmb{\ell'}} \phi_{\mbell-\mbell'} + \mathcal{O}(\phi^2) \,,
\ee
where $W^{\phi^{XY}}_{\mbell, \pmb{\ell'}}$ is the coefficient on the $\mathcal{O}(\phi)$ term.
Information about the lensing potential is thus contained in the off-diagonal elements ($\mbell \ne \mbell'$) of the second term and subsequent higher order terms.
Said another way, lensing distorts the observed CMB anisotropies, which introduces statistical anisotropy in the covariance of the CMB;
these changes are encapsulated in the second and higher-order terms in Equation \ref{eq:lens_expansion}.

The analysis in this paper will proceed in the following manner:
First we filter the CMB fields in order to maximize signal to noise for the lensing calculation.
Next we estimate the two-dimensional lensing potential $\phi$ from these filtered fields.
We correct $\phi$ for two effects: a multiplicative normalization and an additive ``mean-field'' bias correction.
Then we calculate the power spectrum of $\phi$, and subtract two noise-bias terms.
Finally, we calculate the amplitude of the final spectrum relative to a fiducial cosmology.

\subsection{Map Filtering}
\label{sec:cinv}

The first step in this analysis is to filter the CMB fields to maximize the CMB signal relative to the noise.
We use the following data model to motivate our filtering choices.
We will refer to data maps in position-space as $d_j$, where $d \in [T(\hat{\mb{n}}), Q(\hat{\mb{n}}), U(\hat{\mb{n}})]$ and $j$ is the pixel index.
We assume that these data maps are a combination of the sky signal $X_{\mbell} \in [T_{\mbell}, E_{\mbell}, B_{\mbell}]$ and noise given by
\be
d_j = \sum_{\mbell} P_{j \mbell} X_{\mbell} + \sum_{\mbell} P_{j \mbell} N_{\mbell} + n_j \,.
\label{eq:dj}
\ee
Here, $P_{j \mbell}$ is a matrix operator that applies any filter transfer functions (for example, smoothing by the instrumental beam, pixelization, and timestream filtering, e.g., Equation \ref{eq:tf}) and inverse-Fourier transforms the sky signal.
Additionally, $P_{j \mbell}$ transforms $E_{\mbell}$ and $B_{\mbell}$ into $Q(\hat{\mb{n}})$ and $U(\hat{\mb{n}})$.

The map noise is modeled with two components.
The Fourier-domain noise contribution $N_{\mbell}$ represents ``sky noise'' that comes from components of the sky that do not trace the CMB, such as foregrounds or emission from the atmosphere.
The map-domain noise contribution $n_j$ arises from instrumental noise; 
it is assumed to be white and uncorrelated between pixels.
For modes that are heavily filtered, the instrumental noise may be rolled off just as the sky signal is.
We ignore this effect, with the understanding that it causes our inverse-variance filtering procedure to \textit{overestimate} the instrumental variance on such scales (relative to the sky component, which does include a filter-transfer function). 
Over-estimating the instrumental variance in the filter is slightly sub-optimal but will not bias the CMB filtered fields.

We now build a filter tuned to maximize signal-to-noise of CMB anisotropy Fourier modes.
This is essentially a ``matched filter'' technique using an inverse-variance filter for data maps that have anisotropic noise and a filter transfer function.
With a map $d_j$ described by Equation \ref{eq:dj}, we can determine inverse-variance filtered Fourier modes $\bar{X}$ as
\be
\left[ S^{-1} + P^{\dagger} n^{-1} P \right] S \bar{X} = P^{\dagger} n^{-1} d \,.
\ee
We solve for $\bar{X}$, using a conjugate-gradient-descent method to evaluate the expression
\be
\label{eq:xbar_full}
\bar{X} = S^{-1} \left[ S^{-1} + P^{\dagger} n^{-1} P \right]^{-1}  P^{\dagger} n^{-1} d \,,
\ee
where throughout this paper, we will use an over-bar notation for inverse-variance-weighted quantities.
Here $S$ is the total sky signal covariance matrix, defined as the sum of the signal and ``sky noise'' power: $S \equiv C^{X}_{\ell} + C^{N}_{\mbell}$;
in this expression, $C^{X}_{\ell}$ is the theoretical power spectrum of $X$ evaluated on the 2D Fourier plane, 
and $C^{N}_{\mbell}$ is the 2D power spectrum of $N_{\mbell}$ from Equation \ref{eq:dj}, specifically, $\langle|N_{\mbell}|^2\rangle$.
The matrix $n^{-1}$ is the sky mask multiplied by the inverse of the map-noise variance $\langle |n_j^2| \rangle$.
Specifically, $n$ is calculated by taking the map-noise $n_j$ from Equation \ref{eq:dj}
then setting the inverse-noise level  $n_j^{-1} = 0$ for masked pixels $j$ 
(this is equivalent to taking $n_j \rightarrow \infty$ for masked pixels).

The filtered fields $\bar{X}_{\mbell}$ form the input to the estimator of the lensing potential.

\subsection{Quadratic Estimator of $\phi$}
\label{sec:qest}

We now construct an estimator for the lensing potential that takes advantage of the off-diagonal correlations introduced by lensing in the CMB fields \citep{hu02a}.
Calculating a properly normalized, unbiased estimate of $\phi$ requires three steps that are detailed below:
\begin{enumerate}
  \item{Calculate an inverse-variance-weighted estimate of the lensing potential ($\bar{\phi}_{\mbL}$) from two filtered CMB fields,
    i.e., $TT$, $TE$, $EE$, $EB$, and $TB$.}
  \item{Remove a ``mean-field'' bias.}
  \item{Normalize the estimate.  
    In the $\phi$-estimator we use in this paper, the correct normalization of the estimate of $\phi$
    for a given value of $\mbL$. 
    is one over the total inverse-variance weight.
    The best estimate of the lensing potential after normalization will be designated with a hat symbol, $\hat{\phi}_{\mbL}$.}
\end{enumerate}

Using two CMB fields $\bar{X}$ and $\bar{Y}$ that have been filtered as described in Section \ref{sec:cinv}, 
we estimate the inverse-variance-weighted lensing potential as follows:
\be
\label{eq:phi_bar}
\bar{\phi}^{XY}_{\mbL} = \int{d^2\mbell W^{XY}_{\mbell,\mbell-\mbL}}
\bar{X}_{\mbell}\, \bar{Y}^{*}_{\mbell-\mbL}
\ee
where $W^{XY}_{\mbell,\mbell-\mbL}$ is a weight function as described below.
Analogous to $\bar{X}$, we continue to use the over-bar notation to indicate that $\bar{\phi}_{\mbL}$ is an inverse-variance-weighted field.
This estimation technique amounts to taking a weighted sum of the covariance between $\bar{X}_{\mbell}$ and $\bar{Y}_{\pmb{\ell'}}$ at all pairs of angular wavenumbers $\mbell$ and $\pmb{\ell'}$ separated by $\mbell - \pmb{\ell'} = \mbL$. 
As indicated by Equation \ref{eq:lens_expansion}, modes $\mbell$ in $\bar{X}$ and $\pmb{\ell'}$ in $\bar{Y}$ have a covariance that is imprinted by modes in the lensing potential $\phi$ with angular wavenumber $\mbL$.

The weight function $W^{XY}_{\mbell,\mbell-\mbL}$ acts like a matched filter;
in Equation \ref{eq:phi_bar} this weight is convolved in Fourier-space with the two filtered CMB fields to maximize the response to the lensing signal.
We use the minimum-variance estimator for this analysis \citep{hu02a}, in which this weight function is simply the leading-order coefficient on $\phi$ in Equation \ref{eq:lens_expansion}:
$W^{XY}_{\mbell,\mbell-\mbL} \equiv W^{\phi^{XY}}_{\mbell,\mbell-\mbL}$\footnote{
In contrast to \cite{hu02a}, which uses the unlensed CMB spectrum in this definition, while we use the lensed CMB spectrum following the treatment of \cite{lewis11}.}.
In general, other weight functions can be used (for example, ``bias-hardened'' estimators defined in \citealt{namikawa13}).

This estimator is biased since it includes contributions that arise from the statistical anisotropy introduced by non-lensing sources such as the sky mask, in-homogeneous map noise, etc.  
These terms constitute a ``mean-field'' (MF) bias, defined as
\be
\label{eq:phi_MF}
\bar{\phi}^{XY,{\rm MF}}_{\mbL} = \int{d^2\mbell W^{XY}_{\mbell,\mbell-\mbL}}
  \langle \bar{X}_{\mbell}\, \bar{Y}^{*}_{\mbell-\mbL} \rangle \,,
\ee
where the average is taken over realizations of the CMB and noise.
We calculate the MF contribution by averaging over estimates of $\phi$ from Monte-Carlo (MC) simulations with independent realizations of the CMB, lensing potential, and instrument noise.
Since the simulated lensing signal is uncorrelated from simulation to simulation, it averages to zero in this calculation; the common signal that remains after averaging is the MF bias.

Finally, we need to normalize the response of the estimator to produce an unbiased estimate of $\phi$.
This unbiased estimate will be designated with a hat symbol, $\hat{\phi}_{\mbL}$.
We calculate this normalization in a two-step process.
We analytically calculate the normalization as a 2D Fourier-space object,
then correct the analytic normalization using simulations.

Drawing on the intuition that the estimator is a weighted sum of off-diagonal elements in the covariance between $\bar{X}_{\mbell}$ and $\bar{Y}_{\pmb{\ell'}}$, the desired normalization is the reciprocal of the sum of these weights.
To properly calculate the normalization, however, we must account for the map-filtering process described in Section \ref{sec:cinv}.
For the sky signal $X_{\mbell}$ in our maps (specifically, $X_{\mbell}$ as it appears in Equation \ref{eq:dj}),
the C-inverse filter can be described as an operator in two-dimensional Fourier-space that can mix power between $\mbell$-modes.
The C-inverse filter, however, can be approximated as a diagonal Fourier-space filter function of the form:
$[C^{XX}_{\mbell} + C^{NN}_{\mbell}]^{-1} X_{\mbell} \equiv \mathcal{F}^{X}_{\mbell} X_{\mbell}$,    
where $C^{XX}_{\mbell}$ is the power spectrum of field $X_{\mbell}$, 
and $C^{NN}_{\mbell}$ is the power spectrum of the map noise (in which the beam ($F^{\rm beam}_{\mbell}$) and filter-transfer function ($F^{\rm filt}_{\mbell}$) have been divided out in Fourier-space; see Section \ref{sec:data} for more detail).    
We use this diagonal approximation for one thing only: to analytically estimate the normalization of the $\phi$-estimator.

Under this approximation, the normalization is given by
\be
\label{eq:response}
\mathcal{R}^{XY,{\rm Analytic}}_{\mbL} = \int{d^2\mbell\, W^{XY}_{\mbell,\mbell-\mbL}\times W^{\phi^{XY}}_{\mbell,\mbell-\mbL} \mathcal{F}^{X}_{\mb\ell} \mathcal{F}^{Y}_{\mbell-\mbL} } \,.
\ee
Note that in the absence of filtering (i.e., $\mathcal{F}_{\mbell} = 1$), this equation is just the sum of the weights from Equation \ref{eq:phi_bar}.
The approximation $\mathcal{F}^{X}_{\mbell}$ is not exact; 
we therefore calculate a multiplicative normalization correction $\mathcal{R}_{\mbL}^{XY,{\rm MC}}$ from simulations,
\be
\mathcal{R}_{\mbL}^{XY,{\rm MC}} =
        \frac{\langle \hat{\phi}^{I',XY}_{\mbL}\,\, \phi^{I*}_{\mbL}\rangle}
             {\langle \phi^I_{\mbL} \phi^{I*}_{\mbL}\rangle} \,,
\ee
where for each simulated realization, $\phi^I_{\mbL}$ is the input and 
\be
\hat{\phi}^{I',XY}_{\mbL} = \frac{1}{\mathcal{R}^{XY,{\rm Analytic}}_{\mbL}}
  (\bar{\phi}^{XY}_{\mbL} - \bar{\phi}^{XY,{\rm MF}}_{\mbL})
\ee
is the matching reconstruction, normalized by the analytic normalization.
The reconstruction $\hat{\phi}^{I',XY}_{\mbL}$ is masked, the denominator is taken directly from the harmonic-space input maps, 
 and the appropriate factor of $f_{\rm mask}$ (see Equation \ref{eq:clpp_def}) is applied.
The average is taken over 400 simulations.
In principle (or the limit of many, many simulations), the correct normalization would be the product of $\mathcal{R}^{XY,{\rm Analytic}}_{\mbL}$ with $\mathcal{R}^{XY,{\rm MC}}_{\mbL}$ in the 2D $\mbL$-plane,
however, each individual mode in $\mathcal{R}^{XY,{\rm MC}}_{\mbL}$ is still very noisy, making this method intractable.
We expect this normalization to be isotropic, so we average $\mathcal{R}^{XY,{\rm MC}}_{\mbL}$ within annuli in $\mbL$-space to produce 
\be
\mathcal{R}_{L}^{XY,{\rm MC}} = \langle \mathcal{R}_{\mbL}^{XY,{\rm MC}} \rangle \,.
\ee
The normalization we use is the product of the analytically calculated normalization with the annulus-averaged $MC$ correction:
\be
\mathcal{R}_{\mbL}^{XY} \equiv \mathcal{R}_{\mbL}^{XY,{\rm Analytic}} \times \mathcal{R}_{L}^{XY,{\rm MC}} \,.
\ee
We expect the $MC$ correction to be relatively small,
and indeed,  we find that $\mathcal{R}_{L}^{XY,{\rm MC}}$ is a 
$\leq 10\%$ correction.

For each combination of fields $X$ and $Y$, our estimator of the lensing potential is defined as
\be
\label{eq:phi_hat}
\hat{\phi}^{XY}_{\mbL} = \frac{1}{\mathcal{R}^{XY}_{\mbL}}
  (\bar{\phi}^{XY}_{\mbL} - \bar{\phi}^{XY,{\rm MF}}_{\mbL}) \,.
\ee
We calculate all five temperature and polarization estimators $XY \in [TT, TE, EE, EB, TB]$ (a quadratic estimate of $\phi$ from $BB$ has vanishing signal-to-noise in a cosmology with negligible gravitational wave perturbations\footnote{
This can be understood intuitively as follows.
A cosmology with negligible gravitational wave perturbations has negligible primordial B-mode power.
The main contribution of B-mode power will come from E-modes that have been lensed.
This means that when the B-mode power is expanded as Taylor-series in the lensing potential $\phi$, 
the leading-order term will be first-order and higher in $\phi$.
The quadratic $BB$ estimator is constructed from two copies of the B-mode field;
therefore the leading term will be second-order in $\phi$ and will thus be negligible.
} \citealt{lewis06}).

We now want to combine these estimates into one \mv{} estimate of the lensing potential, denoted by $\hat{\phi}^{\rm \mv}_{\mbL}$.
We form $\hat{\phi}^{\rm \mv}_{\mbL}$ from a weighted average of the five estimators,
where the optimal weight $w^{XY}_{\mbL}$ for the \mv{} estimator is $\mathcal{R}^{XY}_{\mbL}$.
This can be understood as follows.
In quadratic maximum likelihood estimators, $\mathcal{R}^{XY}_{\mbL}$ is the Fisher matrix for $\phi^{XY}_{\mbL}$;
this means that the un-normalized estimate $\bar{\phi} - \bar{\phi}^{\rm MF}$ is the inverse-variance-weighted lens reconstruction.
Thus we calculate the average of the inverse-variance-weighted estimators ($\bar{\phi} - \bar{\phi}^{\rm MF}$), and divide by the sum of the normalizations.
This is equivalent to calculating a weighted-sum of the $\hat{\phi}$ estimators, where the weight $w^{XY}_{\mbL}$ is the normalization, $\mathcal{R}^{XY}_{\mbL}$.
For each $\mbL$ mode, 
\be
\label{eq:phi_hat_mv}
\hat{\phi}^{\rm \mv}_{\mbL} = 
  \frac{\sum_{XY} w^{XY}_{\mbL} \hat{\phi}^{XY}_{\mbL} }
       {\sum_{XY} w^{XY}_{\mbL} } \,,
\ee
where the sum is taken over \mbox{$XY \in [TT, TE, ET, EE, EB, BE, TB, BT]$}.
We additionally form a polarization-only (\pp) estimate $\hat{\phi}^{{\rm \pp}}_{\mbL}$ from a weighted average of the $EE$, $EB$, and $BE$ estimators.
Maps and power spectra of the lensing potential are calculated from both the \mv{} and \pp{} estimators.

\subsection{The Power Spectrum of $\phi$}
\label{sec:theory_clpp}

In this section, we present our method for estimating the power spectrum of the lensing potential.
The steps are as follows, with a more detailed description given below.
\begin{enumerate}
  \item{Calculate a cross-spectrum (\clppHatUVXYL) from two estimates of $\phi$, $\hat{\phi}^{UV}_{\mbL}$ and $\hat{\phi}^{XY}_{\mbL}$.}
  \item{Calculate and subtract noise-bias terms ($\left. \Delta C_{\mbL}^{\phi^{UV} \phi^{XY}}\right|_{\rm RDN0}$ and $\left. \Delta C_{\mbL}^{\phi^{UV} \phi^{XY}}\right|_{\rm N1}$ ).  
    The de-biased spectrum for this combination of CMB fields is denoted as \hatclppUVXYL{}, with a hat symbol to indicate that this is our best estimate.}
  \item{Average this spectrum into bins in $L$.  The binned spectrum we report is denoted as $\hat{C}^{\phi^{UV} \phi^{XY}}_{L_b}$.}
  \item{Finally, calculate the amplitude of \hatclppUVXYL{} relative to a fiducial spectrum.
    Note, the amplitude is calculated directly from the 2D $\mbL$-plane, rather than from the binned spectrum.}
\end{enumerate}

The power spectrum from two estimates $\hat{\phi}^{UV}$ and  $\hat{\phi}^{XY}$ is
\be
\label{eq:clpp_def}
 \clppHatUVXYL \equiv 
 f_{\rm mask}^{-1}\langle \hat{\phi}^{UV}_{\mbL} \,\, \hat{\phi}^{*}\,^{XY}_{\mbL} \rangle \, ,
\ee
where $f_{\rm mask}$ is the average value of the fourth power of the apodization and point-source mask.
In the following discussion, we will drop the $UV, XY$ superscripts ($C_L^{\phi \phi} $) unless they are needed for clarity.
Note that in this power spectrum calculation, the mean-field correction $\phi^{\rm MF}$ is estimated from two independent sets of simulations for each of the two $\hat{\phi}$ estimates.
We must do this because our estimates of the mean-field are noisy, owing to a finite number of simulations.
By using separate estimates of the mean-field, we eliminate correlations between the two $\hat{\phi}$ estimates resulting from this noisy subtraction.

\begin{figure}[t]
\begin{center}
\includegraphics[width=\columnwidth]{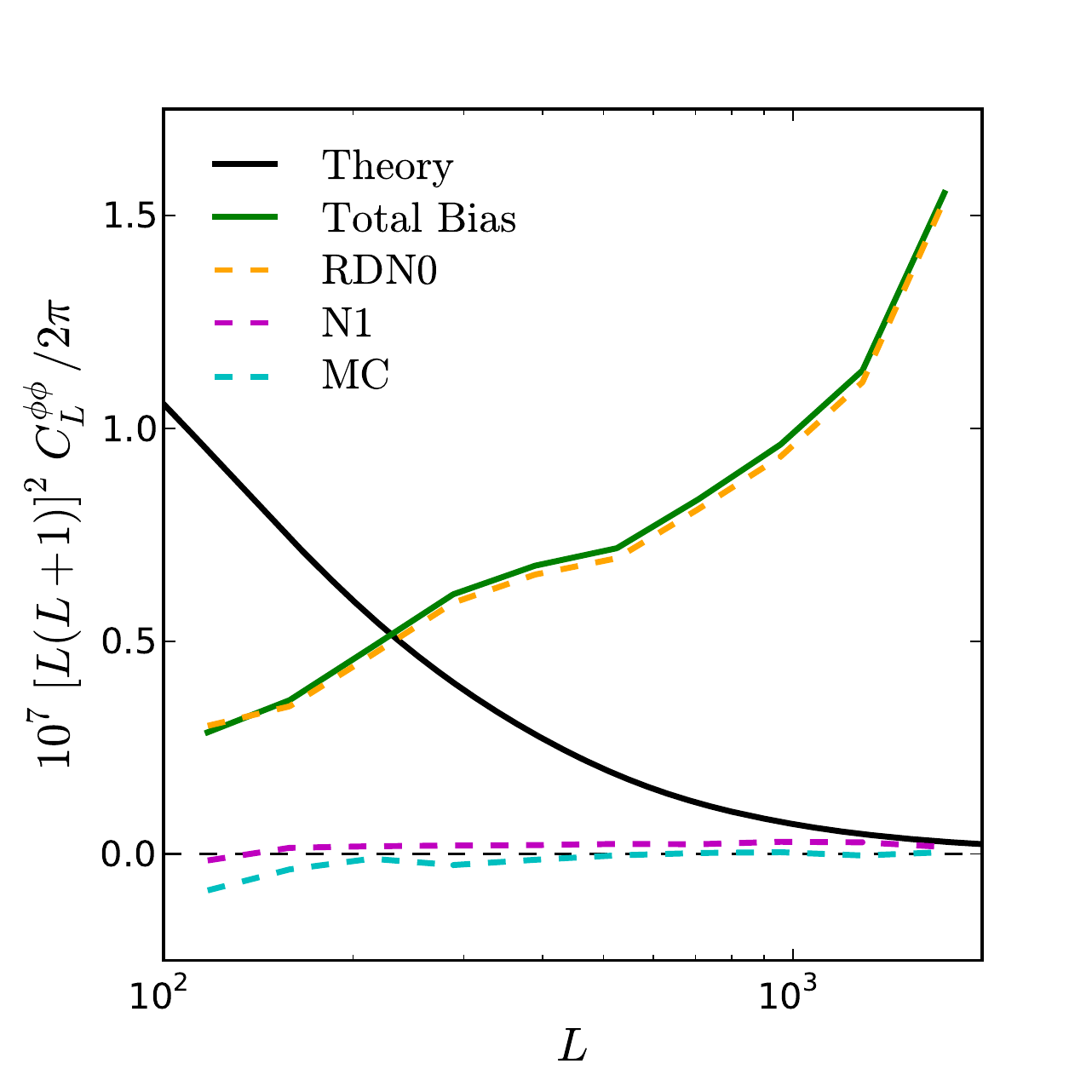}
\end{center}
\caption{
Noise-bias levels for this $\phi$ reconstruction.
The theoretical \LCDM{} lensing potential power spectrum is shown in \textbf{black}, and the individual bias terms defined in Equation \ref{eq:cmbLppHat} are shown individually.
The realization-dependent N0 (RDN0) term (\textbf{dashed orange line}) is used to correct the data for the Gaussian noise bias (see Equation \ref{eq:rdn0}).
The N1 bias (\textbf{dashed purple line}) arises from connected contributions to the CMB trispectrum (see Equation \ref{eq:n1}).
The sum of these two terms is the total bias (\textbf{solid green line}) that we subtract.
The residual Monte-Carlo (MC) bias term (\textbf{dashed cyan line}) from Equation \ref{eq:cmbLppHat} is calculated as the difference between the mean of de-biased lensed simulation spectra and the input spectrum to the simulations.
The total bias shows the reconstruction-noise power in the reconstructed $\phi$ maps; 
thus this reconstruction measures $\phi$ modes with a signal-to-noise ratio greater than 1 
(alternatively, the measurement is $\phi$-sample-variance limited) for $100<L \lesssim 250$.}
\label{fig:bias}
\end{figure}

While this estimate of the \clpp{} power spectrum has minimal variance, it suffers from additive biases which we must subtract.
We now describe these additive bias terms.
The estimates of $\phi$ are quadratic, i.e., 2-point functions of the CMB fields,
thus the cross-spectrum \cmbLppHat{} between estimates of $\phi$ probes the 4-point function (trispectrum) of the CMB.
The CMB trispectrum includes contributions from disconnected and connected pieces, only some of which contain information about \clpp{}.
The remaining correlations show up as ``noise bias'' terms in \cmbLppHat{} \citep{kesden03}.
At a given order, the disconnected pieces arise from lower-order correlations, while the connected pieces are new at each order.
In the case of the CMB trispectrum, the 2-point (Gaussian) correlations in the CMB give rise to the disconnected pieces,
while the higher-order (non-Gaussian) correlations introduced by lensing give rise to the connected pieces.
We model the full cross-spectrum, including bias terms, as:
\be
\cmbLppHat = \hatcmbLpp + \dclNzero + \dclNone + \dclMC
\label{eq:cmbLppHat}
\ee
where \hatcmbLpp{} is the term we want to calculate, \dclNzero{} and \dclNone{} are the disconnected and connected pieces of the trispectrum, respectively, and \dclMC{} encapsulates any remaining bias terms.
We discuss each of these terms below, and show them in Figure \ref{fig:bias}.

The first bias term \dclNzero{} arises from disconnected contributions to the trispectrum.
This term has no dependence on $\phi$ and arises from Gaussian correlations in the CMB fields, foregrounds, and noise; 
this term is called ``N0'' because it is zeroth-order in \clpp.
We estimate the N0 contribution from simulations in the following way:
We create two sets of simulations, $MC$ and $MC'$, with different realizations of the CMB and $\phi$ (including foregrounds and noise).
The prescription for calculating N0 from the $MC$ and $MC'$ simulations can be written as the sum of two terms:
\begin{multline}
\dclNzero = \\
\begin{aligned}
\Big \langle
  &+ \cmbLppHat[\bar{U}_{\rm MC},\bar{V}_{\rm MC'},\bar{X}_{\rm MC},\bar{Y}_{\rm MC'}] \\
  &+ \cmbLppHat[\bar{U}_{\rm MC},\bar{V}_{\rm MC'},\bar{X}_{\rm MC'},\bar{Y}_{\rm MC}]
  \Big \rangle_{\rm MC,MC'}
\end{aligned}
\end{multline}
where we have re-written \clppHatUVXYL{} to explicitly show the dependence on the four input fields $U, V, X, Y$: $\cmbLppHat[\bar{U},\bar{V},\bar{X},\bar{Y}] \equiv \clppHatUVXYL $.

This estimate of the N0 bias for the data will be imperfect because the power spectrum used in the simulations will be different from that of the data.
We can reduce our sensitivity to this difference by using the data itself, a correction called the ``realization-dependent N0'' (RDN0) bias \citep{namikawa13}.
We calculate the RDN0 bias by replacing one of the CMB fields in the estimate of N0 with the data itself, then combining it with the N0 estimate above:
\begin{multline}
\dclRDNzero = \\
\begin{aligned}
\Big \langle
  & +\cmbLppHat[\bar{U}_{\rm d},  \bar{V}_{\rm MC},  \bar{X}_{\rm d},   \bar{Y}_{\rm MC}]
    +\cmbLppHat[\bar{U}_{\rm MC}, \bar{V}_{\rm d},   \bar{X}_{\rm d},   \bar{Y}_{\rm MC}]  \\
  & +\cmbLppHat[\bar{U}_{\rm d},  \bar{V}_{\rm MC},  \bar{X}_{\rm MC},  \bar{Y}_{\rm d} ]
    +\cmbLppHat[\bar{U}_{\rm MC}, \bar{V}_{\rm d},   \bar{X}_{\rm MC},  \bar{Y}_{\rm d} ]  \\
  & -\cmbLppHat[\bar{U}_{\rm MC}, \bar{V}_{\rm MC'}, \bar{X}_{\rm MC},  \bar{Y}_{\rm MC'}] \\
  & -\cmbLppHat[\bar{U}_{\rm MC}, \bar{V}_{\rm MC'}, \bar{X}_{\rm MC'}, \bar{Y}_{\rm MC}] 
  \Big \rangle_{\rm MC,MC'} \,,
\end{aligned}
\label{eq:rdn0}
\end{multline}
where the $d$ subscript indicates a CMB field from the data.
We subtract this RDN0 estimate from the data spectrum.

The second bias term \dclNone{} arises from connected contributions to the trispectrum and depends linearly on \clpp{} \citep{kesden03};
this term is called ``N1'' because it is first-order in \clpp.
We estimate this term by creating two sets of simulations, $MC$ and $MC'$, in which each simulated pair has the same realization of $\phi$, but different realizations of the unlensed CMB;
see Section \ref{sec:sims} for more detail.
The N1 term is then\footnote{
Note, Gaussian foreground power is not included in the simulations for N1.}
\begin{multline}
\label{eq:n1}
\dclNone = \\
\begin{aligned}
\Big \langle
  & +\cmbLppHat[\bar{U}_{\phi^1,{\rm MC}},\bar{V}_{\phi^1,{\rm MC'}},\bar{X}_{\phi^1,{\rm MC}},\bar{Y}_{\phi^1,{\rm MC'}}] \\
  & +\cmbLppHat[\bar{U}_{\phi^1,{\rm MC}},\bar{V}_{\phi^1,{\rm MC'}},\bar{X}_{\phi^1,{\rm MC'}},\bar{Y}_{\phi^1,{\rm MC}}] \\
  & -\dclNzero
  \Big \rangle_{{\rm MC,MC'}}
\end{aligned}
\end{multline}

The final term \dclMC{} encapsulates any corrections that have not been accounted for as yet.
We calculate this term as the difference of the simulations from the input spectrum:
\be
\dclMC = \cmbLppHat{}_{\rm MC} - \dclNzero - \dclNone
\ee
We find that \dclMC{} is small and do not subtract it in our final estimate; see Section \ref{sec:errors} for further discussion.

Thus our final measured power spectrum may be written as
\be
\hatcmbLpp = \cmbLppHat - \dclRDNzero - \dclNone \,.
\ee

Now that we have the best estimate of our lensing power spectrum \hatcmbLpp{}, we would like to average this spectrum into bins in $L$.
The power in each bin is referred to as a ``bandpower.''
Because \hatcmbLpp{} is a two-dimensional quantity with varying signal-to-noise across the 2D $\mbL$-plane, a naive binning operation of simply averaging \hatcmbLpp{} in each bin is sub-optimal.
Instead, we calculate the weighted average of \hatcmbLpp{} within each bin:
\be
C^{\phi^{UV} \phi^{XY}}_b \equiv \frac{\sum_{\mbL \in b} w^{UV XY}_{\mbL} \hatclppUVXYL }
                                  {\sum_{\mbL \in b} w^{UV XY}_{\mbL} } \,.
\ee
The choice of the weight function $w^{UV XY}_{\mbL}$ follows from the same reasoning used to define the weights in Equation \ref{eq:phi_hat_mv}.
Specifically, in the quadratic maximum likelihood estimator that we are using, $\mathcal{R}^{XY}_{\mbL}$ is the Fisher matrix for $\phi^{XY}_{\mbL}$; 
therefore the inverse-variance weight for \hatcmbLpp{} is the product of the normalization of the two estimates $\phi^{UV}$ and $\phi^{XY}$:
\be
\label{eq:amp_weight}
w^{UV XY}_{\mbL} = \mathcal{R}_{\mbL}^{UV} \mathcal{R}_{\mbL}^{XY} \,.
\ee
This weight scheme is analytically optimal in two cases: 
either for an estimator of the auto-spectrum where $U=V=X=Y$ or for a complete set of cross-spectra between different estimators (e.g., the \mv{} and \pp{} estimators).

We calculate the amplitude $A_b$ of our data spectrum relative to a theory spectrum within a bin $b$,
\be
\label{eq:amp_exact}
A^{UV XY}_b \equiv \frac{C^{\phi^{UV} \phi^{XY}}_b }
           {C^{\phi^{UV} \phi^{XY}, {\rm theory}}_b } \,,
\ee
where the binned theory spectrum is
\be
\label{eq:clth_b}
C^{\phi^{UV} \phi^{XY}, {\rm theory}}_b \equiv \frac{\sum_{\mbL \in b} w^{UV XY}_{\mbL} C^{\phi \phi, {\rm theory}}_{\mbL}}
                                  {\sum_{\mbL \in b} w^{UV XY}_{\mbL} } \,.
\ee
We report bandpowers $\hat{C}^{\phi \phi}_{L_b}$ as the data-amplitude $A_b$ multiplied by 
the theoretical spectrum $C^{\phi \phi, {\rm theory}}_{L_b}$ evaluated at the bin center $L_b$, 
\be
\label{eq:bandpowers}
\hat{C}^{\phi \phi}_{L_b} \equiv A_b C^{\phi \phi, {\rm theory}}_{L_b} \,.
\ee

Finally, we calculate two overall amplitudes of \hatcmbLpp{} relative to a theory spectrum:
$A_{\rm \mv}$ for the \mv{} spectrum and $A_{\rm \pp}$ for the polarization-only spectrum.
These amplitudes are calculated using Equation \ref{eq:amp_exact}, where the ``bin'' is taken to the full range $100< L <2000$.
The reference theory spectrum $C^{\phi \phi, {\rm theory}}_{L}$ is taken to be the \textsc{Planck+Lens+WP+highL} spectrum. 
We additionally report the spectrum bandpowers as defined in Equation \ref{eq:bandpowers}, which contain the information about the shape of our measured spectrum.

\section{Simulations}
\label{sec:sims}

The lensing analysis presented here relies heavily on accurate simulations.
We use the spectra from the \textsc{Planck+Lens+WP+highL} model as our fiducial cosmological model for simulations.
We create simulations as follows.

First, we create realizations of spherical-harmonic coefficients ($a_{lm}$) for the CMB fields $T$, $E$, and $B$ (including the proper correlations between the fields), as well as the lensing potential $\phi$.
We simulate modes with $\ell < 6000$.
We evaluate the spherical harmonic transform \footnote{
The spherical harmonic transform is performed with routines from the HEALPIX library \citep{gorski05}.}
of the $a_{lm}$ coefficients on a grid with an equidistant cylindrical projection (ECP) \citep{lewis05}.
The lensing operation is applied by distorting the unlensed fields using the deflection map derived from $\phi$.
These distorted maps are interpolated back onto a fixed ECP grid, creating lensed CMB skies with full non-Gaussian information.

These maps are then mock-observed with the map-making pipeline.
This involves creating simulated TOD for each bolometer, filtering the TOD, 
accumulating the TOD into flat-sky maps using the SPT pointing information, and coadding different detectors based on the individual detector weights.
In effect, the resulting maps are what SPT would have seen if these simulated skies had been the true CMB sky, in the absence of noise and foregrounds.

The mock-observed maps are transformed back into Fourier space, where Gaussian foreground power is added.
We use the foreground model from \cite{story13}, with the following components:
$D^{\rm PS}_{3000}=10 \muksq$ is the power from Poisson-distributed point-sources that scales as $D^{\rm PS} \propto \ell^2$,
$D^{\rm CL}_{3000}=5 \muksq$ is the power from clustered CIB sources that scales as $D^{\rm CL} \propto \ell^{0.8}$,
$D^{\rm SZ}_{3000}=5 \muksq$ is the amplitude of the tSZ power spectrum that we use to scale the thermal SZ template taken from \cite{shaw10}.
The three coefficients here are given at $\ell=3000$.
See Section 6.1 of \cite{story13} for details.
The Fourier-space maps are multiplied by the Fourier-space SPT beam $F^{\rm beam}_{\mbell}$.

We then add realizations of noise to the simulations.
These realizations are estimated directly from the data.
We take all observations and divide them into two halves, and subtract the coadd of one half from the coadd of the other half.
We create many realizations of the noise by choosing random sets of halves.
This method of calculating the noise variance gives an unbiased but potentially noisy estimate; 
 however, given the number of independent observations used in this work, the noise on the variance estimate is expected to be negligible.

We make three sets of simulations:
\begin{enumerate}
  \item{Set ``A'': 500 lensed simulations.}
  \item{Set ``B'': 100 lensed simulations with different realizations of the CMB but the \textit{same} realizations of $\phi$ as the first 100 simulations in Set A.}
  \item{Set ``C'': 500 unlensed simulations.}
\end{enumerate}

These simulations are used as follows.
The simulations in set ``A'' are used to calculate 
the mean-field (see Section \ref{sec:qest}), 
the N0 bias term (see Section \ref{sec:theory_clpp}), 
and the statistical uncertainty (see Section \ref{sec:errors}).
Specifically, the first 100 simulations are used to calculate the mean-field; 
in each cross-spectrum we use the first 50 simulations to estimate the mean-field of the first $\phi$ and the second 50 simulations to estimate the mean-field of the second $\phi$.
The remaining 400 simulations are used to calculate the statistical uncertainty on the lensing spectrum and amplitude.
The entire set is used to calculate \dclRDNzero.

The simulations in set ``B'' are used to calculate \dclNone{}, as described in Equation \ref{eq:n1}.
Specifically, the simulations labeled $MC$ are comprised of 50 simulations from set ``A'', while the simulations labeled $MC'$ come from 50 matching simulations from set ``B''.

Finally, the unlensed simulations that comprise set ``C'' are used for two purposes.
First, we check that no spurious lensing in our pipeline is detected by measuring the amplitude of the reconstructed lensing potential from these unlensed simulations.
The average amplitude and variance of these simulations is $A_{\rm unl} = -0.024 \pm 0.065$, thus passing this test.
Second, we quantify how significantly we reject the no-lensing hypothesis by comparing the lensing amplitude measured in the data to the variance of these unlensed simulations.
The first 100 simulations are used to calculate the mean-field, and the remaining 400 simulations are used to calculate the statistical uncertainty of unlensed skies.

In Figure \ref{fig:amp_hist}, the distribution of lensing amplitudes from these lensed simulations is shown in green, 
and the distribution for unlensed simulations is shown in red.

\section{Uncertainty budget}
\label{sec:errors}

\begin{figure*}[t]
\begin{center}
\begin{minipage}{0.435\linewidth}
\begin{subfigure}{}
\includegraphics[width=\textwidth]{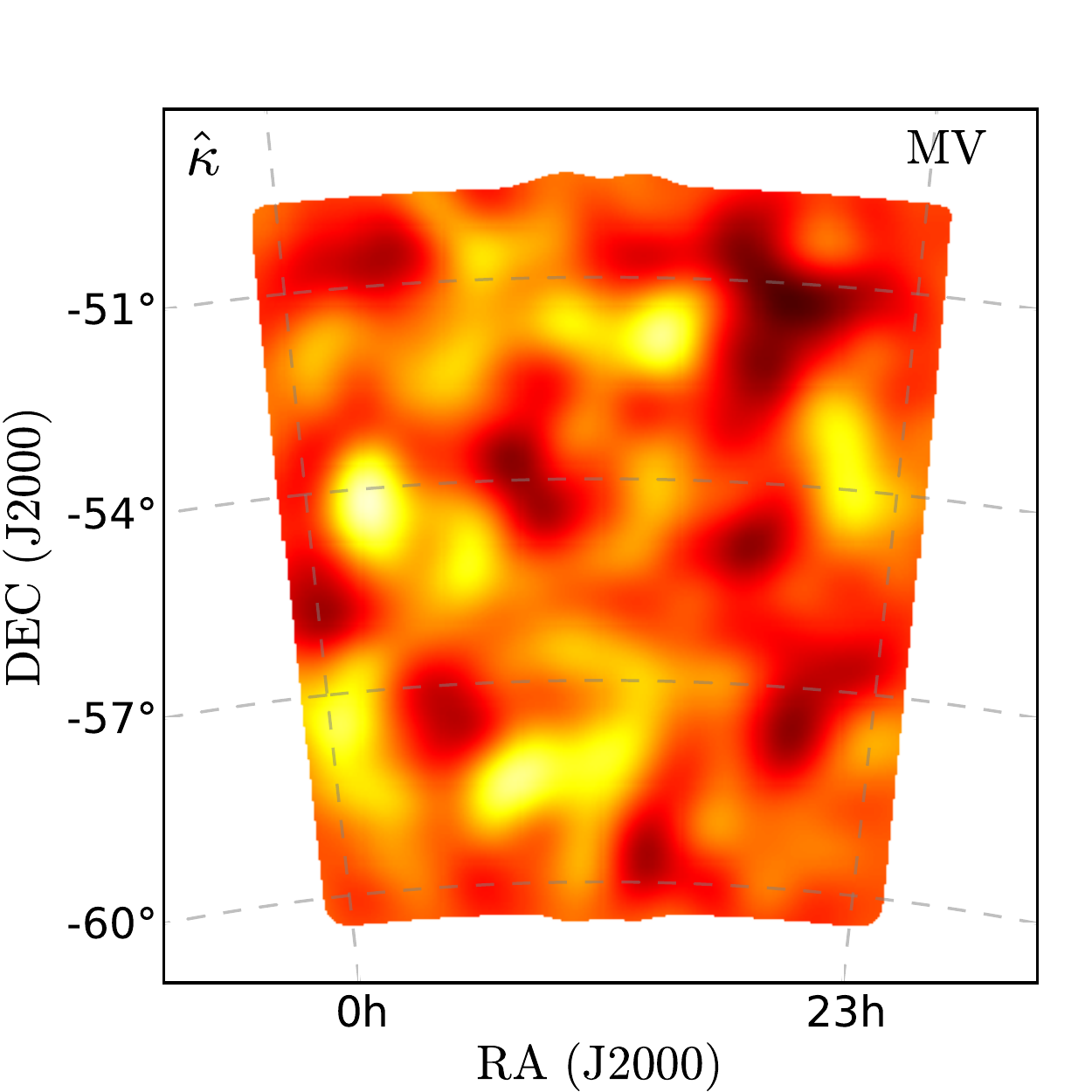}
\end{subfigure}
\end{minipage}
\begin{minipage}{0.40\linewidth}
\begin{subfigure}{}
\vspace{-0.5cm}
\\ 
\includegraphics[trim=20 15 15 15, clip, width=0.45\textwidth]{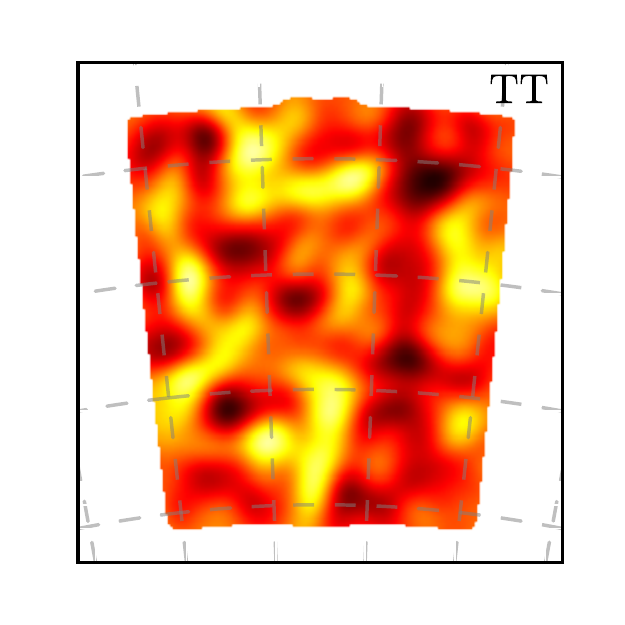}
\includegraphics[trim=20 15 15 15, clip, width=0.45\textwidth]{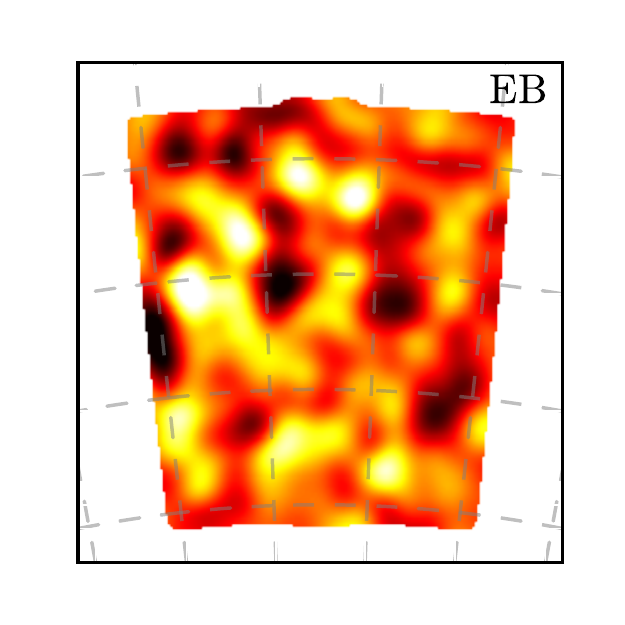} 
\\
\includegraphics[trim=20 15 15 15, clip, width=0.45\textwidth]{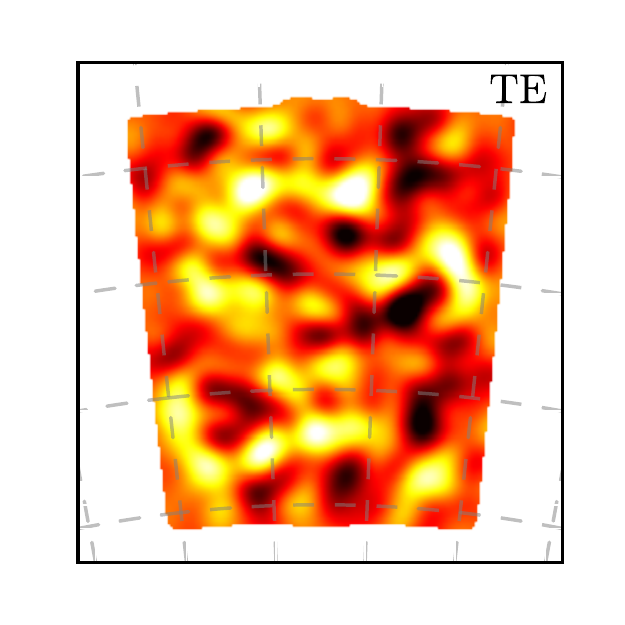}
\includegraphics[trim=20 15 15 15, clip, width=0.45\textwidth]{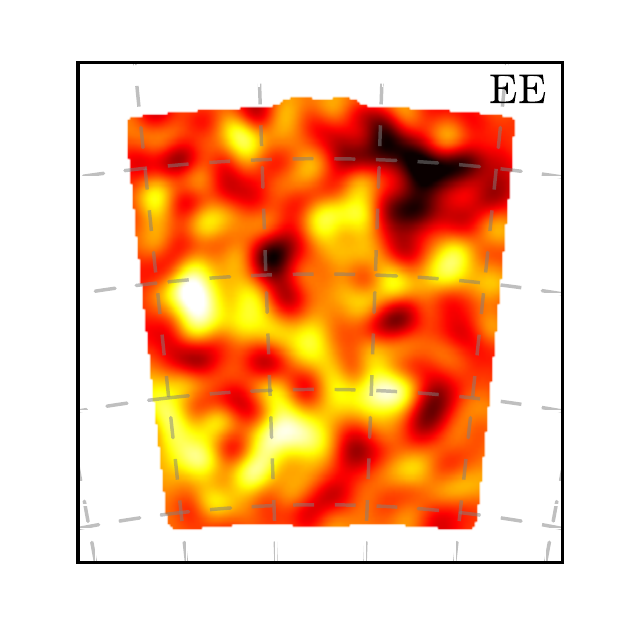}
\end{subfigure}
\end{minipage}
\begin{minipage}{0.1\linewidth}
\includegraphics[trim=5 0 120 10, clip, height=3.5\linewidth]{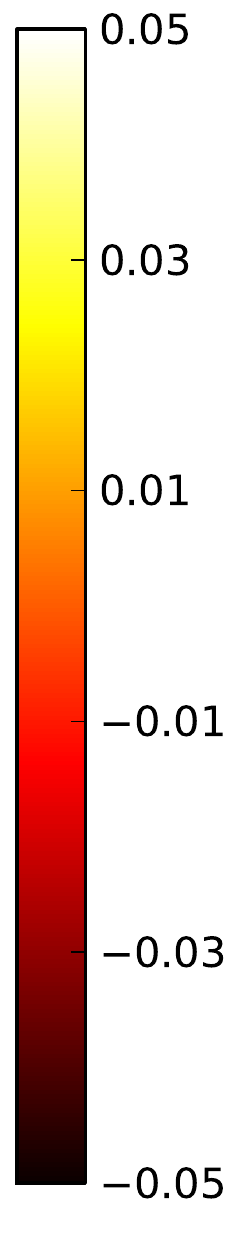}
\end{minipage}
\vspace{-0.2in}
\caption{
Lensing $\kappa$ maps reconstructed from the \sptpol{} 100 deg$^2$ deep-field data, smoothed with a 1-degree Gaussian beam.
The colorbar on the far right shows the color scale, which has been fixed for all $\kappa$ maps in Figures \ref{fig:kappa_map} and \ref{fig:kappa_sims}.
\textbf{Left:} The $\kappa$-map for our \mv\ lensing estimator, which combines all temperature and polarization information.
\textbf{Right}: Individual $\kappa$ estimates from the TT, EB, TE, and EE estimators, with the same color scale.}
\label{fig:kappa_map}
\end{center}
\end{figure*}

\begin{figure*}[t]
\begin{center}
\begin{subfigure}{}
\includegraphics[width=0.3\textwidth]{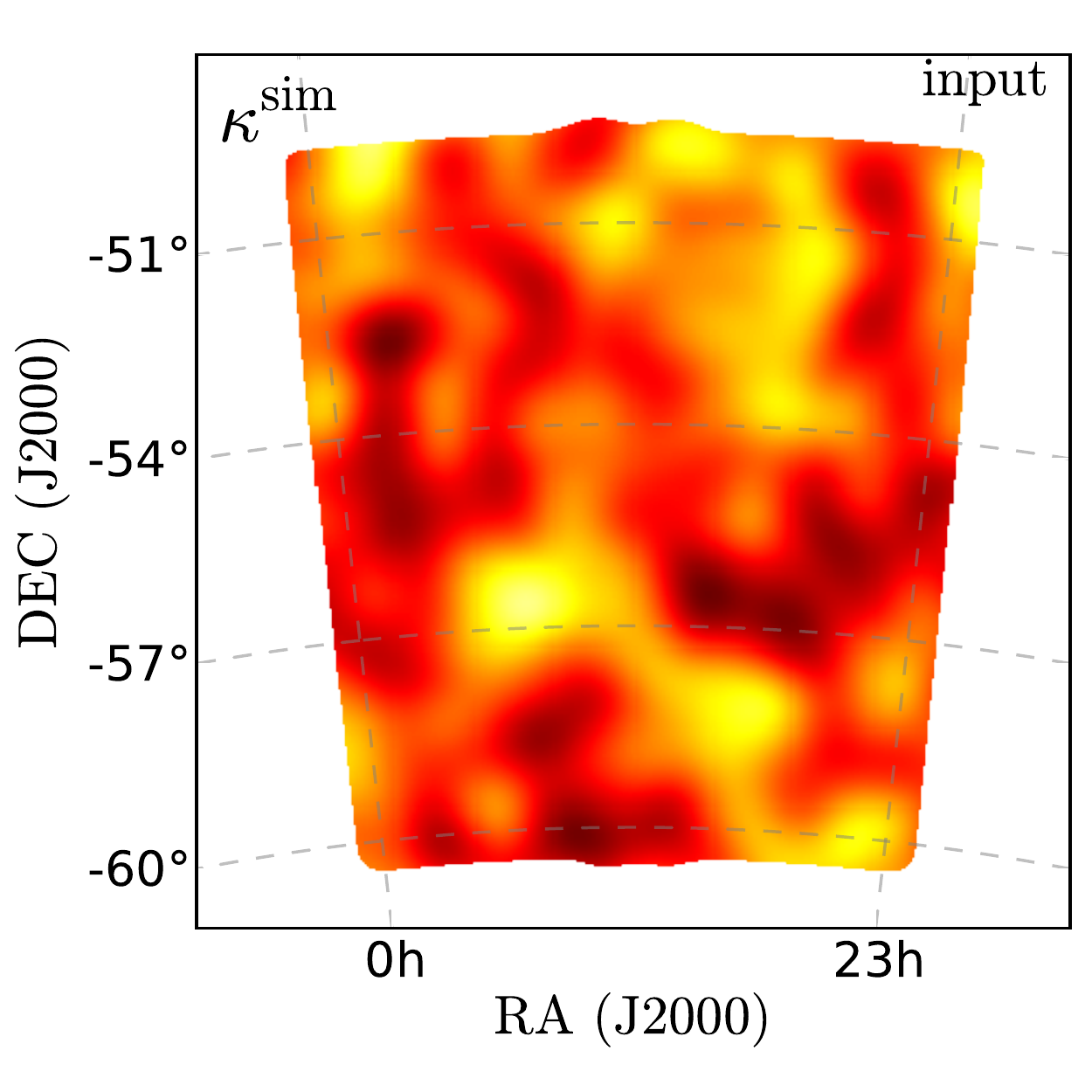}
\end{subfigure}
\begin{subfigure}{}
\includegraphics[width=0.3\textwidth]{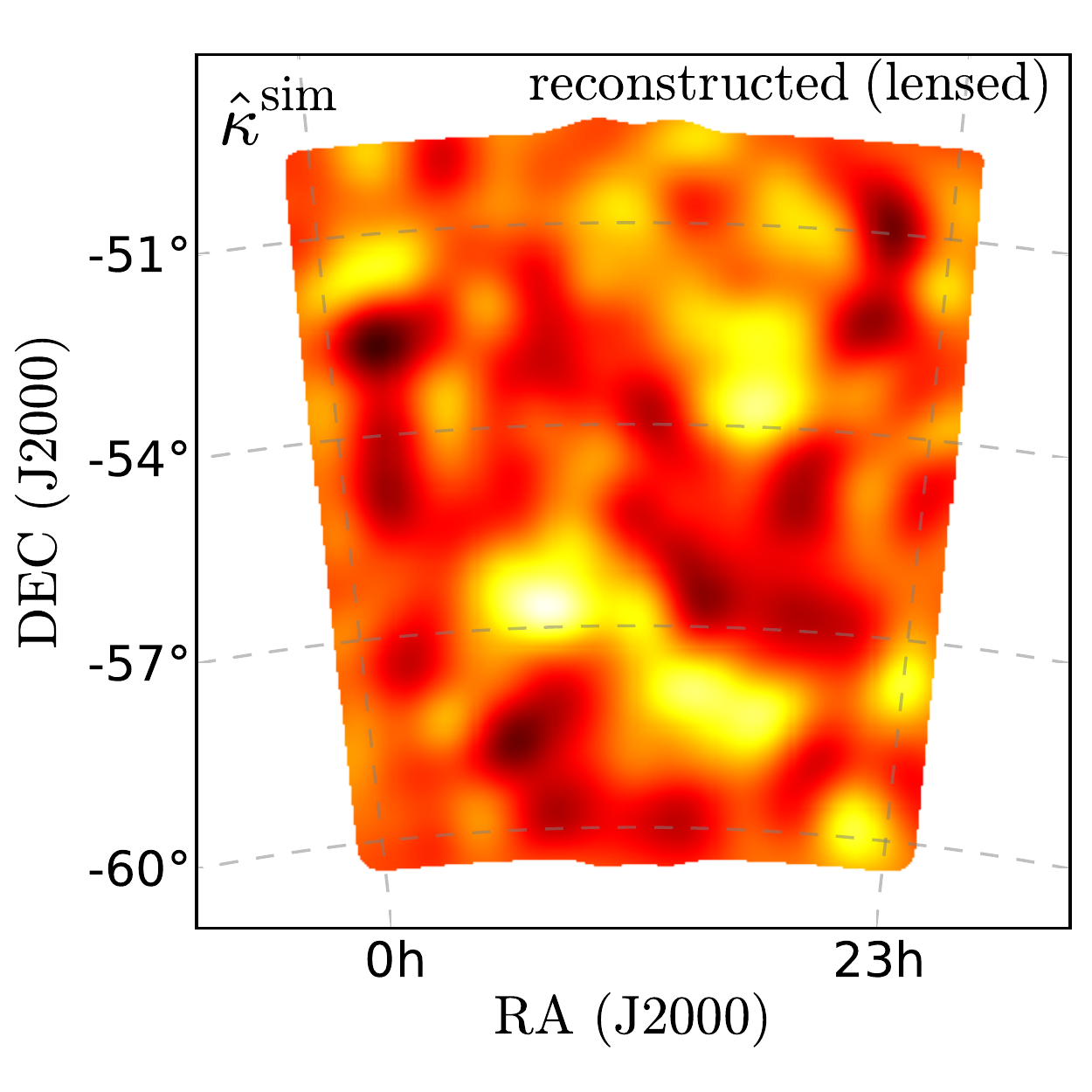}
\end{subfigure}
\begin{subfigure}{}
\includegraphics[width=0.3\textwidth]{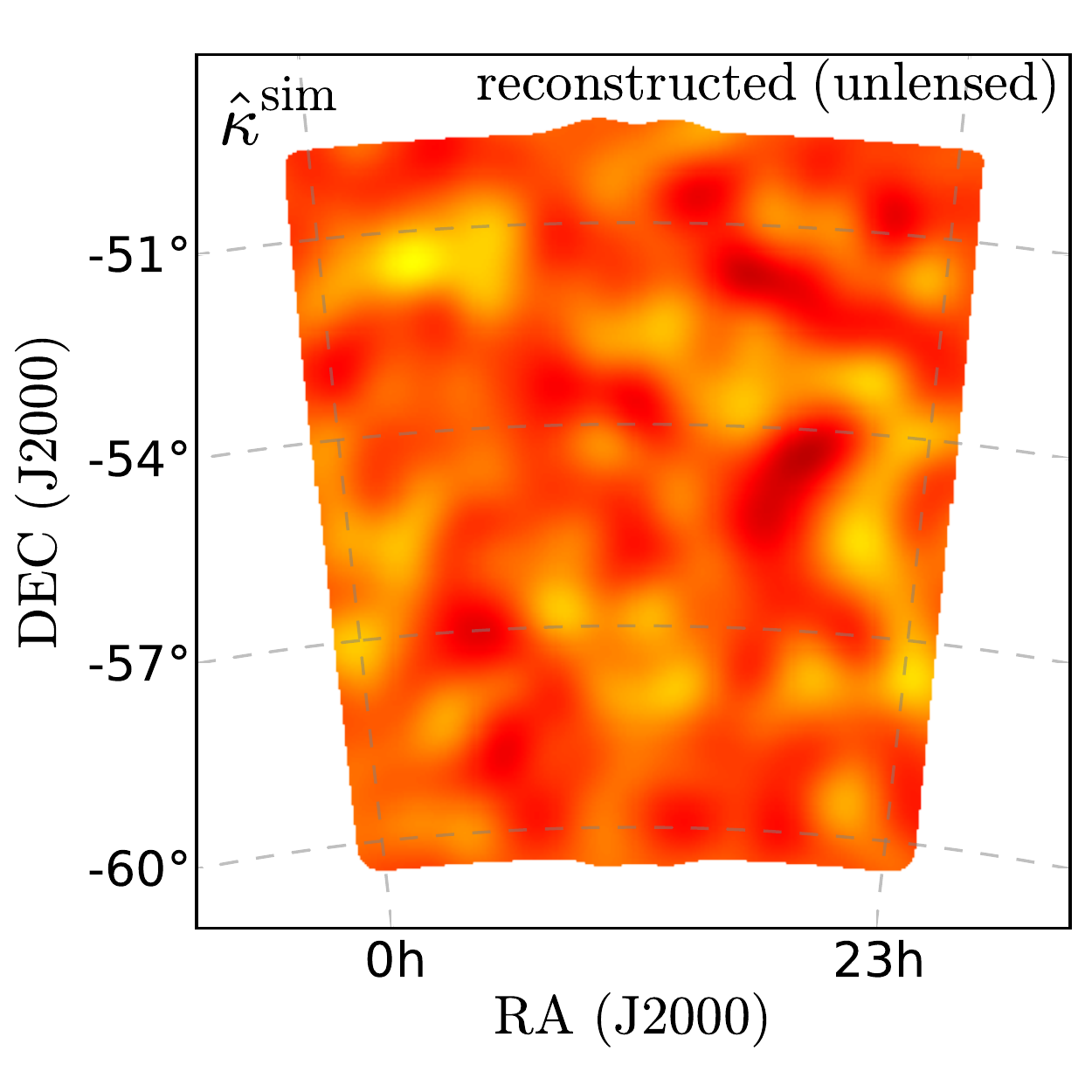}
\end{subfigure}
\caption{Example simulated $\kappa$-maps, plotted with the same color scale as Figure \ref{fig:kappa_map}.
  \textbf{Left:} a simulated input $\kappa$-map.
  \textbf{Middle:} the reconstructed $\kappa$-map estimated from a noisy simulation that has been lensed using the potential shown in the left panel.
  \textbf{Right:} the reconstructed $\kappa$-map estimated from an unlensed simulation.  
  Comparing the reconstructed lensed $\kappa$-map to the input map gives a visual sense of the fidelity of this reconstruction,
  and comparing to the unlensed $\kappa$-map gives a sense of the signal-to-noise in the \mv{} $\kappa$-map.
}
\label{fig:kappa_sims}
\end{center}
\end{figure*}

\begin{table}
\caption{Systematic Uncertainties}
\centering
\begin{tabular}{l | c | c }
\hline\hline
Type & $\Delta A_{\rm \mv}$ & $\Delta A_{\rm \pp}$ \\ [0.5ex]
\hline
$\Delta A_{\rm Tcal}$      & \mvSysTcal  & \ppSysTcal \\
$\Delta A_{\rm Pcal}$      & \mvSysPcal  & \ppSysPcal \\
$\Delta A_{\rm crosstalk}$ & \mvSysXtalk & \ppSysXtalk \\
$\Delta A_{\rm amp}$       & \mvSysAmp   & \ppSysAmp \\
\hline
$\Delta A_{\rm tot}$       & \mvAmpSys  & \ppAmpSys
\end{tabular}
\tablecomments{The contributions to the systematic uncertainty budget.
  The quadrature sum of the systematic uncertainty terms gives the total systematic uncertainty, $\Delta A_{\rm tot}$.}
\label{tab:errors}
\end{table}

The uncertainties on the lensing spectrum and amplitude are comprised of variance (noise and sample), calibration and beam error, and other systematic errors.
The variance component is calculated as the covariance between the power spectra of 400 lensed simulations.
These simulations use independent realizations of the unlensed CMB, lensing potential, and instrumental noise,
thus this procedure naturally accounts for both instrumental noise and sample variance.

We also account for sources of systematic uncertainty.
The resulting uncertainty on the lensing amplitude is calculated below and shown in Table \ref{tab:errors}.

\subsection{Beam and Absolute Calibration}
The estimation of the beam and absolute calibration contributes additional systematic uncertainty.
The fractional uncertainties on the beam measurements are less than $1\%$ over the multipoles used in this analysis, and the resulting uncertainty on the lensing amplitude is negligible.
The $1\sigma$ uncertainly on the absolute calibration in temperature is $\delta_{\rm Tcal}=1.3\%$ (C14).
This will propagate into an uncertainty on the spectrum amplitude of $\Delta A_{\rm Tcal}=A (1+\delta_{\rm Tcal})^4$.
We thus incorporate a systematic uncertainty on the amplitude from calibration and beams of 
$\Delta A_{\rm Tcal}=\mvSysTcal$.

\subsection{Polarization Calibration}
Similarly, an error in the polarization calibration will propagate to an uncertainty on the spectrum amplitude of polarized estimators.
We estimate the $1\sigma$ uncertainty on the polarization calibration to be 
$\delta_{\rm Pcal} = 1.7\%$ (C14).
We incorporate a systematic uncertainty on the \pp{} amplitude of 
$\Delta A_{\rm \pp}^{\rm Pcal}=\ppSysPcal$.
We estimate the resulting systematic uncertainty in the \mv{} amplitude by calculating the relative change in the analytic normalization $\mathcal{R}^{POL,{\rm Analytic}}_{\mbL}$ resulting from a $1\sigma$ shift in $P_{\rm cal}$.
We find an uncertainty in the \mv{} amplitude of 
$\Delta A_{\rm \mv}^{\rm Pcal}=\mvSysPcal$.

\subsection{Temperature-to-Polarization Leakage}
Mis-estimating the temperature power leaked into $Q$ and $U$ maps would bias the lensing measurement.
We recalculate the \mv{} lensing amplitude without correcting for the leakage.
The amplitude changes by less than $0.07\sigma$; this source of uncertainty is negligible.

\subsection{Electrical crosstalk between detectors}
There is some low-level electrical crosstalk between detectors. 
The main effect of crosstalk is to introduce a difference between the temperature and polarization instrumental beams, resulting in a small multiplicative bias in the polarization measurements; see C14 for a detailed discussion.
The effect of crosstalk on this lensing analysis was investigated using the simulation pipeline described in Section \ref{sec:sims} and adding the effect of crosstalk to the simulated timestreams.
We find that crosstalk introduces a small bias in the lensing amplitude, which we estimate to be 
a $5\%$ ($6\%$) in the \mv{} (\pp{}) amplitude;
we account for this by adding a contribution to our systematics uncertainty budget of 
$\Delta A^{\rm crosstalk}_{\mv} = \mvSysXtalk$ and $\Delta A^{\rm crosstalk}_{\pp} = \ppSysXtalk$.
The bias in the data from crosstalk is expected to be smaller than this estimate, because the data uses an RDN0 bias subtraction that reduces the effect of a mismatch between the power in the simulations relative to the data itself.
In contrast, these simulations use a normal N0 bias subtraction, rather than recalculating an RDN0 bias for each simulation.
We keep this over-estimate of the crosstalk-induced bias and include this term in our systematic uncertainty budget.

\subsection{Foregrounds}
Foreground emission from extra-galactic sources and galactic dust contributes both Gaussian power and non-Gaussian signal to CMB observations which, if not accounted for, will bias lensing reconstruction measurements.
The Gaussian power component contributes to the N0 bias, which we subtract using the simulations described in Section \ref{sec:sims} and procedure described in Section \ref{sec:theory_clpp}.

The non-Gaussian mode-coupling from foreground emission has been studied in detail in \cite[hereafter V14]{vanengelen14a}, and \citet{osborne14}.
The work in V14 is particularly relevant to our analysis.
They studied a comprehensive list of potential biases to the temperature lensing reconstruction that arise from foregrounds;
they considered Poisson-distributed galaxies, CIB emission from clustered galaxies, tSZ signal from galaxy clusters, galaxy-lensing correlations, and tSZ-lensing correlations.
For the point-source and cluster masking thresholds used in our analysis, the bias to the lensing spectrum never exceeds a few percent.
\cite{vanengelen12} also tested diffuse Galactic cirrus emission and found that the bias was less than 2\% in all $L$-bins.

The contribution to the N0 bias from polarized foregrounds is negligible at the sensitivity level of this analysis
since the polarized power from point sources is too low to be detected significantly in the $EE$ and $TE$ polarization spectra in C14, even with the significantly higher flux cut of $50$ mJy in that work.
The non-Gaussian signature of polarized foregrounds is expected to be negligible as well:
the polarization fraction of foreground emission is expected to be lower than the polarization fraction of the CMB,
implying that the ratio of non-Gaussian foreground signal to CMB lensing signal will be lower in polarization than temperature,
and the non-Gaussian contribution in temperature has already been shown to be negligible for this analysis.

Finally, we consider whether emission from polarized Galactic dust could contaminate the polarized lensing estimators.
The Gaussian power in $B$ modes from Galactic dust within the \bicep{} field was estimated in a joint analysis of \bicep{} and \planck{} data \citep{bicep2keckplanck15};
 because the \bicep{} field contains the \sptpol{} deep field, this analysis provides a good estimate of Galactic dust in the \sptpol{} data.
The best-fit model from the joint analysis is 
 $D^{{\rm dust}}(\ell) = D^{{\rm dust}}_{\ell=80} \times (\ell/80)^{-0.42}$, where $D^{{\rm dust}}_{\ell=80} = 0.0118$ \muKarcmin.
Using this model, the dust power in this field is 
 $\sim 10\%$ of the lensed $B$-mode power and 
 $\sim 1\%$ of the $B$-mode noise power (as calculated in \citet{keisler15})
 at the lowest multipoles used in the \sptpol{} lensing analysis ($\ell>450$), and the dust power drops rapidly with increasing $\ell$.
Thus we conclude that the Gaussian power from Galactic dust in our polarized maps has a negligible contribution to both our observed signal and variance.
Significant non-Gaussian contributions from any foreground sources, including Galactic dust, should cause a failure of the curl test since foregrounds should contain both gradient and curl modes.
Both the \mv{} and \pp{} estimators pass the curl test, thus we conclude that Galactic dust does not contribute significantly to the observed lensing signal in either temperature or polarization. 

We conclude that given the level of precision of this analysis and our level of source masking, 
the RDN0 bias correction sufficiently accounts for the Gaussian power from unpolarized foregrounds, 
and that the Gaussian power from polarized foregrounds can be neglected in the RDN0 term. 
Furthermore, we conclude that the biases due to the non-Gaussian contribution from polarized and unpolarized foregrounds are negligible at the level of statistical precision of the current analysis.

\subsection{Normalization Calculation}
We find that the mean \mv{} (\pp{}) amplitude of our lensed simulations is 3\% (4.5\%) below unity; this can be seen as a shift in the histogram of lensed simulations in Figure \ref{fig:amp_hist}.
This bias could be an additive term (e.g., the \dclMC{} term from Equation \ref{eq:cmbLppHat}) 
or a multiplicative term (e.g., mis-estimation of the power-spectrum normalization).
We treat this potential bias as a systematic uncertainty in our analysis 
of $\Delta A^{\rm amp}_{\rm \mv}=+\mvSysAmp$ and $\Delta A^{\rm amp}_{\rm \pp}=+\ppSysAmp$
on the \mv{} and \pp{} amplitudes, respectively.

\subsection{Total Uncertainty}
The four sources of significant systematic error described above are added in quadrature to calculate our final systematic uncertainty on the lensing amplitude:
$\Delta A_{\rm \mv}=\pm \mvAmpSys$ and 
$\Delta A_{\rm \pp}=\pm \ppAmpSys$
on the amplitudes of the \mv{} and \pp{} spectra, respectively.
Thus our uncertainty on the amplitude of the \mv{} lensing power spectrum is 
$\Delta A_{\rm \mv}=\pm \mvAmpStat {\rm\, (Stat.)} \pm \mvAmpSys {\rm\, (Sys.)}$,
and our uncertainty on the amplitude of the \pp{} lensing power spectrum is
$\Delta A_{\rm \pp}=\pm \ppAmpStat {\rm\, (Stat.)} \pm \ppAmpSys {\rm\, (Sys.)}$.

Finally, we calculate a total uncertainty by adding the statistical and systematic uncertainties in quadrature:
$\Delta A^{\rm tot}_{\rm \mv}=\pm \mvDampTot$ and 
$\Delta A^{\rm tot}_{\rm \pp}=\pm \ppDampTot$.

\section{Results}
\label{sec:results}

We present three main results in this section:
the \mv{} map of the lensing convergence field, $\kappa(\nhat)$,
the \mv{} and \pp{} estimates of the binned lensing potential power spectrum,
and the amplitudes of these two spectra relative to the \textsc{Planck+Lens+WP+highL} model.

\subsection{Lensing Potential Map}
\label{sec:kappa_map}

We show measured lensing convergence maps\footnote{
The map is presented as $\kappa$ because this corresponds visually to the density of the integrated matter field along the line of sight:
$\kappa>0$ corresponds to an over-density that ``stretches'' the observed CMB pattern,
while $\kappa<0$ corresponds to an under-density that ``contracts'' the observed CMB pattern.}
$\kappa(\nhat)$ in Figure \ref{fig:kappa_map}.
These $\kappa$-maps are the real-space equivalent of $(1/2)L(L+1)\hat{\phi}_{\mbL}$, 
where $\hat{\phi}_{\mbL}$ is defined in Equation \ref{eq:phi_hat}.
We show the \mv{} map, as well as four of the five individual estimators.
The similarities between the different estimators are visually apparent.

We measure lensing modes with a signal-to-noise ratio greater than one for modes between $100<L<250$;
thus the large-scale features that are visible by eye are real over-densities and under-densities in the projected mass distribution in the universe.
This is the highest signal-to-noise lensing map from the CMB to date.

\subsection{Lensing Potential Power Spectrum}
\label{sec:clpp}

\begin{figure*}[t]
\begin{center}
\includegraphics[width=1.0\textwidth]{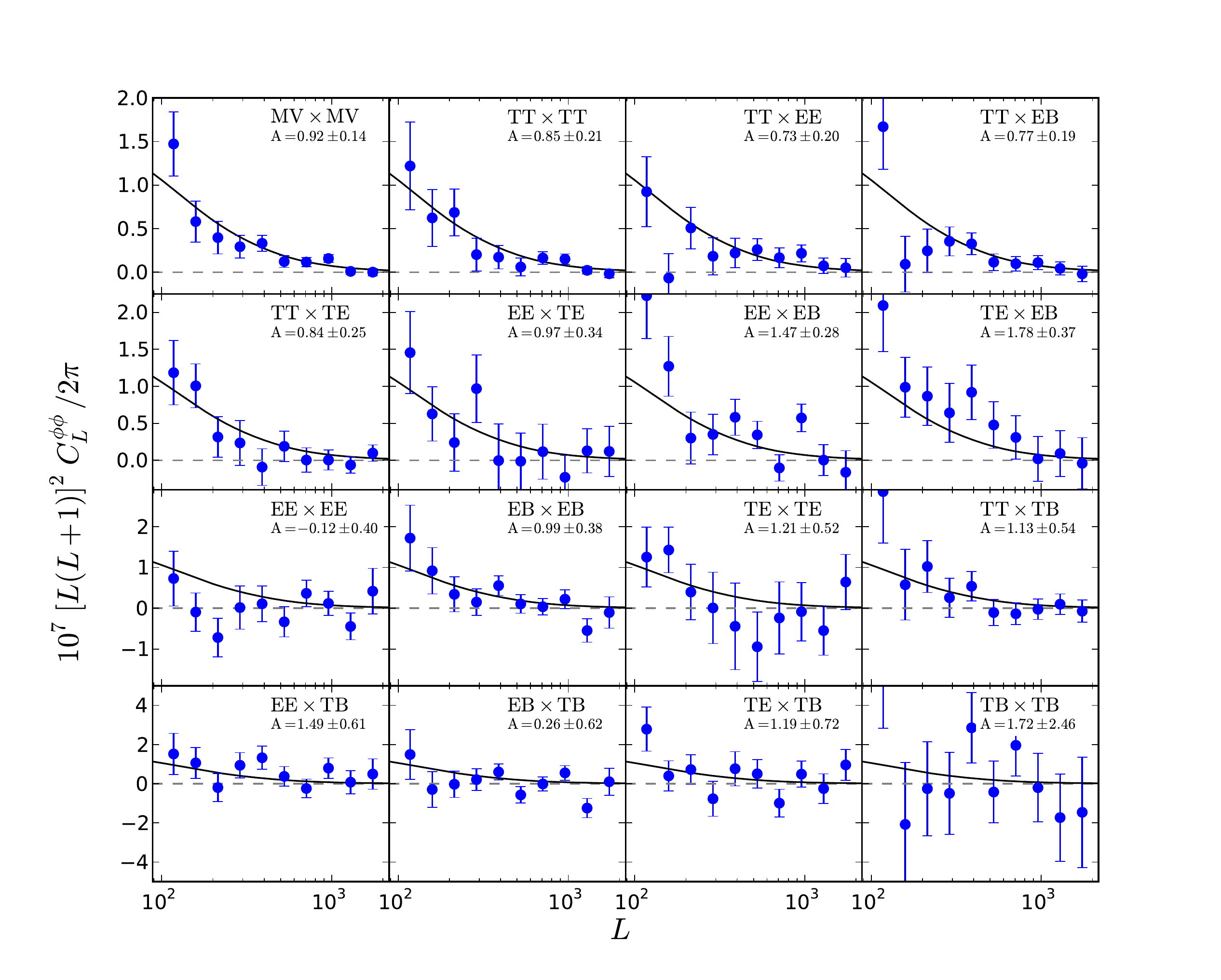}
\caption{The $\hat{C}^{\phi \phi}_{L_b}$ power spectra for all estimators we consider in this paper.  
  ``\mv{} $\times$ \mv{}'' is the spectrum from the minimum-variance estimator.
  The amplitudes of each spectrum relative to the fiducial \textsc{Planck+Lens+WP+highL} model are calculated with Equation \ref{eq:amp_exact} and shown in each panel.
}
\label{fig:clpp_16panel}
\end{center}
\end{figure*}

\begin{figure*}[t]
\begin{center}
\includegraphics[width=0.9\textwidth]{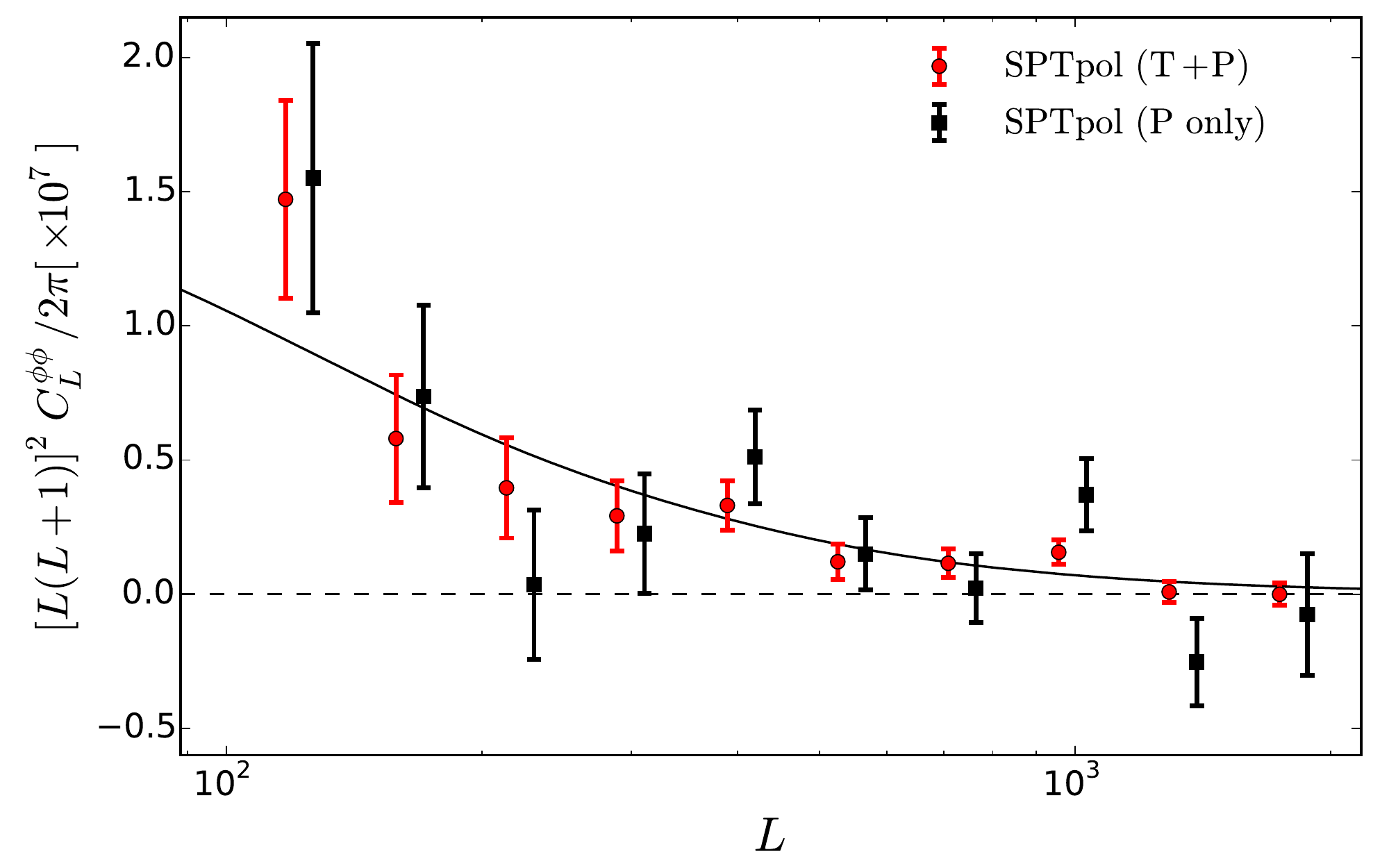}
\end{center}
\caption{Lensing potential power spectrum bandpowers estimated from \sptpol.
The \mv{} and \pp{} spectra are shown with red circles and black squares, respectively.
The black solid line shows the \textsc{Planck+Lens+WP+highL} best-fit \LCDM{} model.
Note the \pp{} points have been shifted by $1/4$ of a bin in $L$ for plotting purposes.}
\label{fig:bandpowers_sptOnly}
\end{figure*}

\begin{figure*}[t]
\begin{center}
\includegraphics[width=0.9\textwidth]{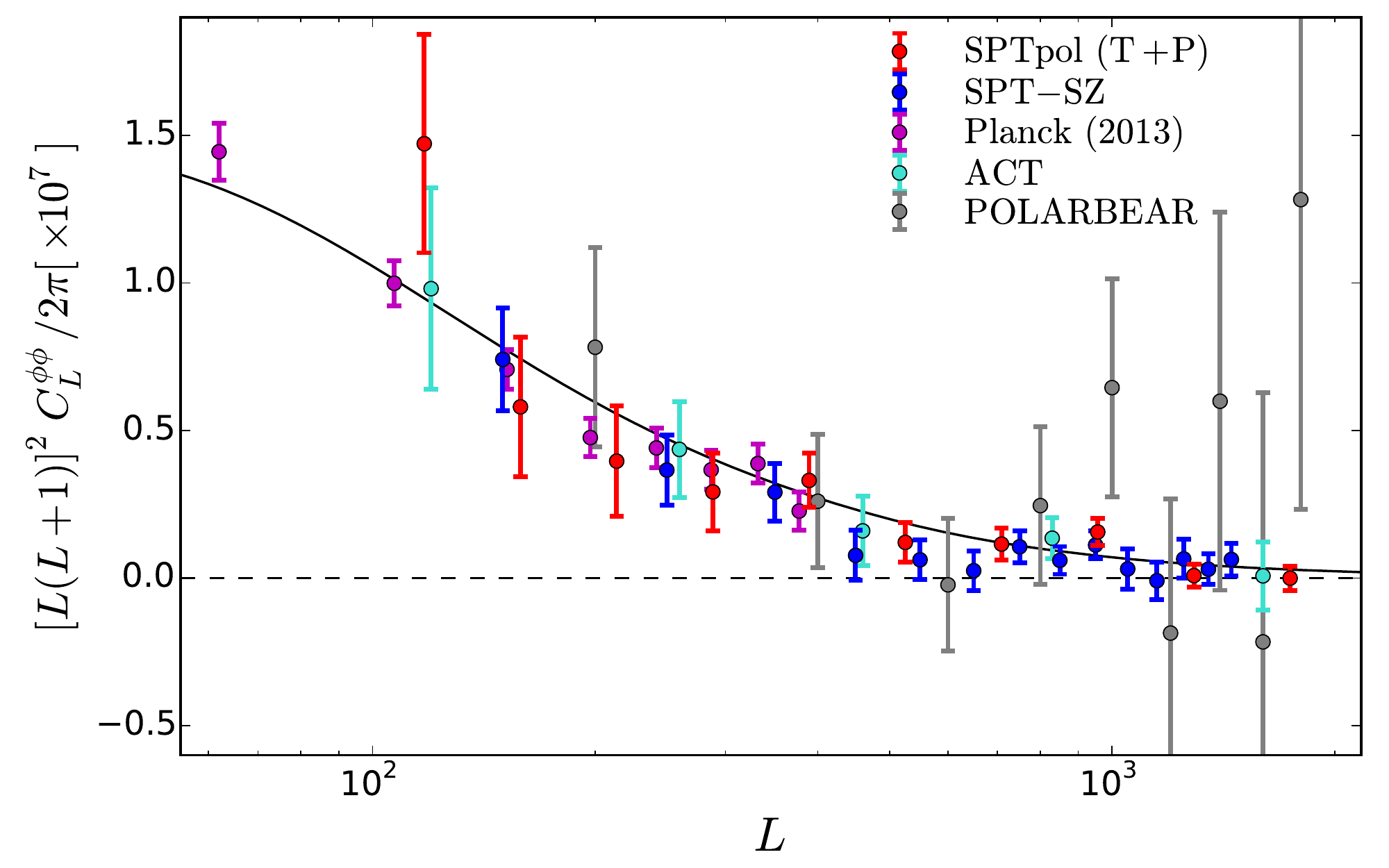}
\end{center}
\caption{Lensing potential power spectrum bandpowers estimated from \sptpol, as well as those previously reported for temperature by
SPT-SZ \citep{vanengelen12},
ACT \citep{das14},
\planck{} \citep{planck13-17},
and for polarization by \polarbear{} (\citeauthor{polarbear2014a}).
The black solid line shows the \textsc{Planck+Lens+WP+highL} best-fit \LCDM{} model.}
\label{fig:bandpowers_summary}
\end{figure*}

Here we present the bandpowers from our estimate of the lensing potential power spectrum.
We report the bandpowers between $100 < L < 2000$.
The lower boundary is set at 100 to maximize the signal while keeping the $MC$ bias sub-dominant to other systematic uncertainties; 
 the $MC$ bias increases as this lower boundary is reduced.
The upper boundary is deep in the noise-dominated region of the spectrum; 
 we find that the statistical uncertainty on the lensing spectrum amplitude is roughly constant for any boundary value above $L\gtrsim1000$ and $L\gtrsim800$ for the \mv{} and \pp{} estimators, respectively.
We calculate the cross-spectrum from each pair of estimators [$\hat{\phi}^{UV}_{\mbL}$, $\hat{\phi}^{XY}_{\mbL}$], 
as well as two minimum-variance estimates, $\hat{C}^{\rm \mv}_{L_b}$ from the combination of all estimators, and $\hat{C}^{\rm \pp}_{L_b}$ from the polarization-only estimators $EE$, $EB$, and $BE$.
The cross-spectra between each pair of these lensing potential estimators is shown in Figure \ref{fig:clpp_16panel}.

The \mv{} spectrum $\hat{C}^{\rm \mv}_{L_b}$ and \pp{} spectrum $\hat{C}^{\rm \pp}_{L_b}$ are shown in Figure \ref{fig:bandpowers_sptOnly}.
The bandpowers for the \mv{} spectrum are presented in Table \ref{tab:bandpowers}.
We measure the amplitude of the spectrum using Equation \ref{eq:amp_exact} relative to the \textsc{Planck+Lens+WP+highL} model.
We find an amplitude of 
$A_{\rm \mv}=\mvAmp \pm \mvAmpStat {\rm\, (Stat.)} \pm \mvAmpSys {\rm\, (Sys.)}$
for the combined spectrum
and $A_{\rm \pp}=\ppAmp \pm \ppAmpStat {\rm\, (Stat.)} \pm \ppAmpSys {\rm\, (Sys.)}$
for the polarization-only spectrum.
Thus we measure the amplitude of the lensing potential power spectrum using solely polarization estimators with a precision of \ppPercent{};
including systematic errors, this becomes a precision of \ppTotPercent{}.
Similarly, the precision of the \mv{} measurement of the amplitude is \mvPercent{};
including systematic errors, this becomes a precision of \mvTotPercent{}.

We quantify how significantly we reject the null hypothesis of no lensing by comparing the data to unlensed simulations.
In 400 unlensed simulations, none have a lensing amplitude as large as that in the data in either the \mv{} or \pp{} estimator.
To estimate statistical significance, we fit a Gaussian to the 400 unlensed simulations, 
yielding a \mvSigUnl{} and \ppSigUnl{} rejection of the null hypothesis for the \mv{} and \pp{} estimators, respectively.

\begin{figure}[t]
\begin{center}
\includegraphics[width=\columnwidth]{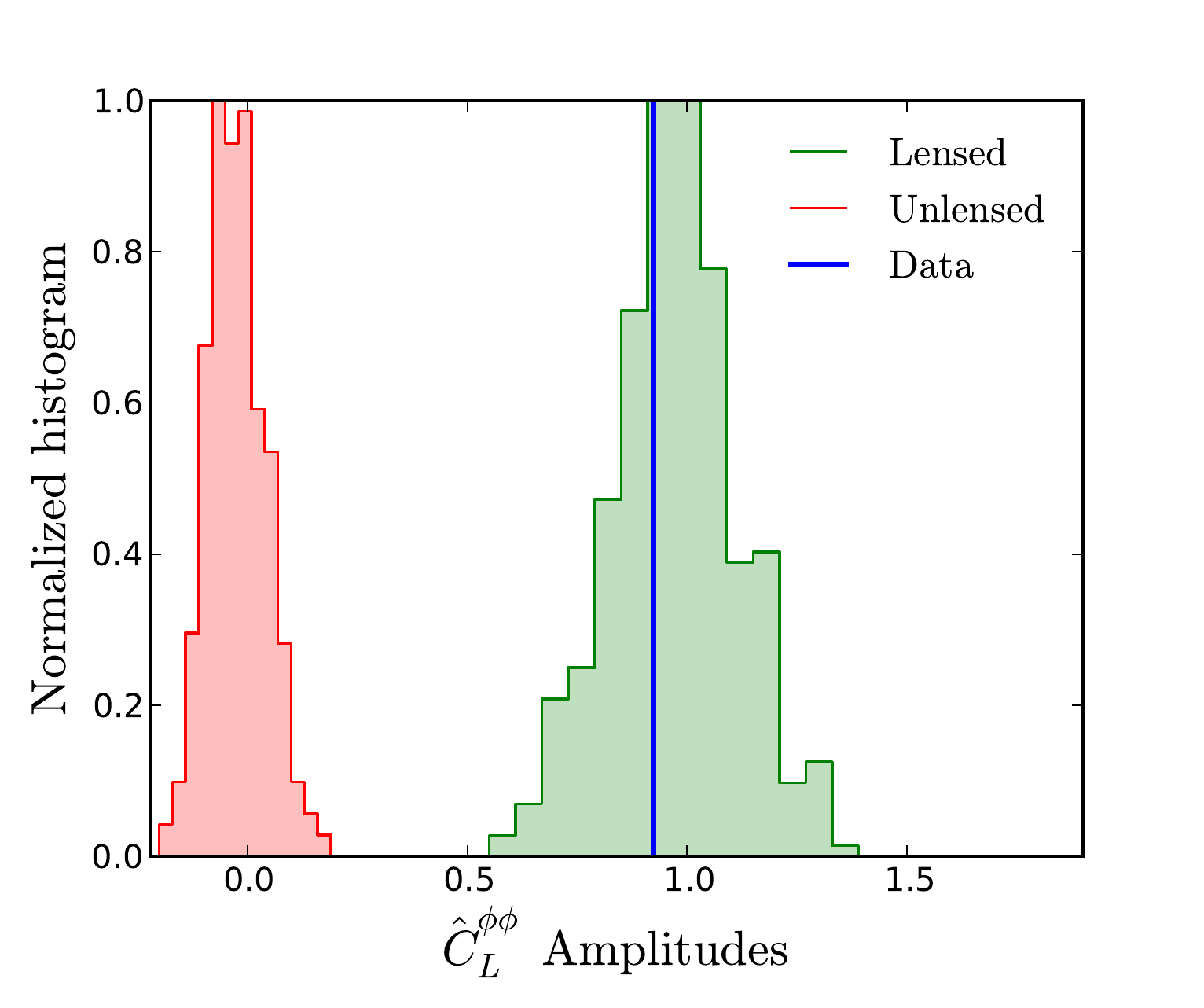}
\end{center}
\caption{
The distribution of reconstructed \mv{} lensing amplitudes from simulations are shown here for lensed (green) and unlensed (red) simulations.
The amplitude of the \mv{} estimate for the data is shown as a blue line.
The statistical uncertainty of the \mv{} lensing construction is given by the standard deviation of the lensed simulations ($\Delta A_{\rm \mv}=\mvAmpStat$).
The significance with which we rule out the no-lensing hypothesis is calculated from the standard deviation of the unlensed simulations (\mvAmpUnlStat).
}
\label{fig:amp_hist}
\end{figure}

Figure \ref{fig:amp_hist} demonstrates the difference between these constraints.
We rule out the no-lensing hypothesis much more significantly than the precision with which we measure the amplitude of the lensing spectrum (\mvSigUnl{} vs. \mvPercent{} $\sim$ \mvSig).
This difference indicates that we measure the presence of lensing modes very significantly, 
however, the measurements are sample-variance dominated on large scales, 
limiting the precision of our constraint on the lensing power spectrum amplitude.

We plot the \mv{} spectrum in comparison with other measurements in Figure \ref{fig:bandpowers_summary}.
All spectra are visually consistent at the current sensitivity levels; 
we discuss these measurements in Section \ref{sec:discussion}.
We compare our \mv{} spectrum amplitude directly to the most precise of these measurements (\planck) 
by recalculating the \mv{} amplitude relative to the same fiducial spectrum used in \cite{planck13-17}.
\planck{} measured an amplitude of $A_{\rm \planck}=0.94 \pm 0.04$;
we find an amplitude relative to the same cosmology of $A_{\rm \mv}^{\rm \planck-Fiducial} = 0.90 \pm \mvDampTot$, which is consistent.

\begin{table}
\caption{\mv\ lensing bandpowers}
\centering
\begin{tabular}{c c c | c }
\hline\hline
$[\,L_{\rm min}$ & $L_{\rm max}\,]$ & $L_b$ & $10^7 [L_b(L_b+1)]^2 \hat{C}_b^{\phi\phi} / 2\pi$ \\ [0.5ex]
\hline
$[\,100$&$133\,]$&$117$&$ \phantom{-}1.47\pm0.37$\\
$[\,134$&$181\,]$&$158$&$ \phantom{-}0.58\pm0.24$\\
$[\,182$&$244\,]$&$213$&$ \phantom{-}0.40\pm0.19$\\
$[\,245$&$330\,]$&$288$&$ \phantom{-}0.29\pm0.13$\\
$[\,331$&$446\,]$&$389$&$ \phantom{-}0.331\pm0.092$\\
$[\,447$&$602\,]$&$525$&$ \phantom{-}0.121\pm0.067$\\
$[\,603$&$813\,]$&$708$&$ \phantom{-}0.115\pm0.053$\\
$[\,814$&$1097\,]$&$956$&$ \phantom{-}0.156\pm0.046$\\
$[\,1098$&$1481\,]$&$1290$&$ \phantom{-}0.008\pm0.038$\\
$[\,1482$&$1998\,]$&$1741$&$-0.000\pm0.041$\\
\hline
\end{tabular}
\tablecomments{The bandpowers for the \mv{} spectrum are presented here as defined in Equation \ref{eq:bandpowers} 
  and shown in Figures \ref{fig:bandpowers_sptOnly} and \ref{fig:bandpowers_summary}.
Bins are evenly spaced in $\log(L)$, and bandpowers are reported at the center of each bin.}
\label{tab:bandpowers}
\end{table}

\section{Systematic Error Checks}
\label{sec:systematics}

\begin{figure}
\begin{center}
\includegraphics[width=\columnwidth]{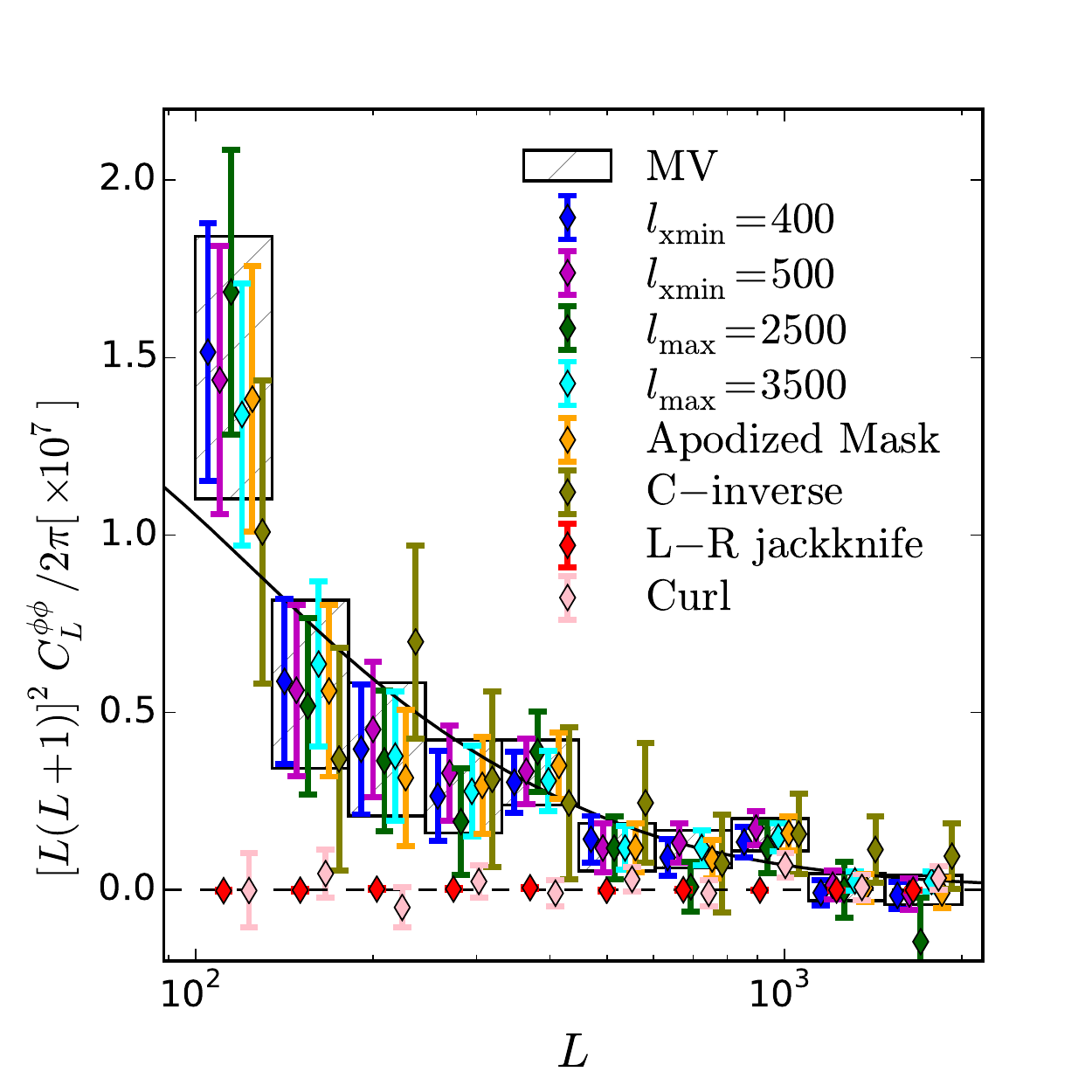}
\includegraphics[width=\columnwidth]{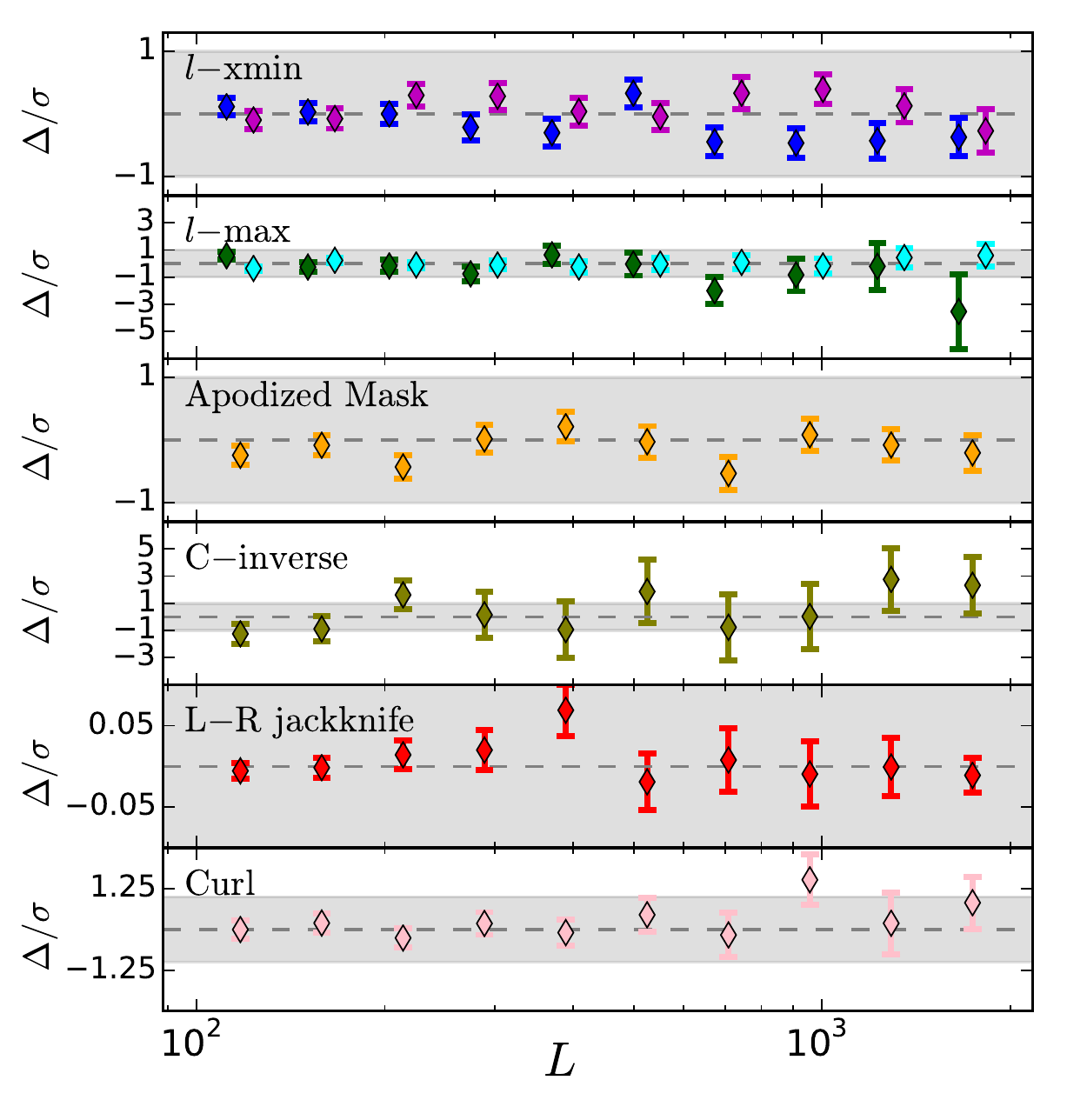}
\end{center}
\label{fig:systematics}
\caption{
Power spectrum consistency tests. 
The \textbf{upper panel} shows the spectrum of each consistency test.
The \textbf{lower panels} show the same set of consistency tests, plotted as the difference between each consistency test relative to the fiducial spectrum, divided by the $1\sigma$ statistical error bar of the \mv{} spectrum.
Note that the statistical uncertainty on these difference-spectra will be a function of the analysis and not necessarily the same as for the baseline \mv{} spectrum.
The grey band in each of the lower panels shows the $1\sigma{}$ statistical uncertainty region.
The error bars shown in the lower panels are calculated from the variance in simulations of each consistency test.}
\end{figure}

\begin{table}
\caption{\mv\ Systematic Error and Null tests}
\centering
\begin{tabular}{l |c c|c c}
\hline\hline
Test Name & $\chi^2$ & (PTE)  & $\Delta$ $A_{\rm \mv}$ & (PTE) \\ 
&&& $\pm {\rm var}(\Delta A_{\rm MV, sim})$&\\ [0.5ex]
\hline
  \rule{0pt}{4ex}
  L-R jackknife        &   7.1 & (0.72) &  0.0040 $\pm$ 0.0052 & (0.45) \\ 
  $l_{\rm xmin} = 400$ &  17.5 & (0.06) & -0.063 $\pm$ 0.030 & (0.025) \\ 
  $l_{\rm xmin} = 500$ &  10.7 & (0.38) &  0.053 $\pm$ 0.033 & (0.10) \\ 
  $l_{\rm max} = 2500$ &  13.6 & (0.19) & -0.122 $\pm$ 0.107 & (0.26) \\ 
  $l_{\rm max} = 3500$ &   9.2 & (0.51) &  0.007 $\pm$ 0.067 & (0.91) \\ 
  Apodized Mask        &  13.6 & (0.19) & -0.043 $\pm$ 0.034 & (0.22) \\ 
  C-inverse            &   9.6 & (0.47) &  0.146 $\pm$ 0.534 & (0.78) \\ 
  Curl                 &   7.4 & (0.69) &  0.082 $\pm$ 0.092 & (0.39) \\ 

\hline
\end{tabular}
\tablecomments{Results of systematics tests.  
  For each test, the $\chi^2$ and PTE of the \clpp{} spectrum are shown in the second column.
  The change in amplitude and associated PTE are shown in the third column.
  See Equation \ref{eqn:sys} for more detail.
  The $l_{\rm xmin} = 400$ test fails, which is why we place the cut higher, at $l_{\rm xmin} = 450$ in the analysis.
}
\label{tab:systematics}
\vspace{0.1cm}
\end{table}

We perform a suite of tests for systematic errors in our data.
For each test, we change one aspect of the analysis and recalculate the lensing spectrum ($C^{\phi \phi}_b\ _{\rm sys}$) and amplitude ($A_{\rm \mv, sys}$).
We then calculate the difference between this spectrum (or amplitude) and the spectrum (or amplitude) from the baseline analysis
(the baseline spectrum and amplitude are the \mv{} results we report in Section \ref{sec:results}).
The difference-spectrum $\Delta C^{\phi \phi}_b$ and difference-amplitude $\Delta A_{\rm \mv}$ 
are the shifts induced in the spectrum and the amplitude, respectively, by the systematic under consideration;
they can be expressed as 
\ba
\label{eqn:sys}
\Delta C^{\phi \phi}_b &=& C^{\phi \phi}_{b,\, {\rm sys}} - C^{\phi \phi}_b \\
\Delta A_{\rm \mv}     &=& A_{\rm \mv, sys} - A_{\rm \mv} \,.
\ea
The difference-spectra and difference-amplitudes are calculated for the data as well as for each simulation.

For each test, we use two metrics to determine if the data pass the test;
these metrics are shown in Table \ref{tab:systematics}.
The first metric considers shifts in the lensing spectrum.
We calculate the $\chi^2$ of the data difference-spectrum ($\Delta C^{\phi \phi}_{b,\,{\rm data}}$) 
using the variance of the simulation difference-spectra ($\sigma_{b,{\rm sys}}$) as the uncertainty.
This can be expressed as 
\be
\chi^2_{\rm sys} = \sum_b{ \frac{(\Delta C^{\phi \phi}_{b,\,{\rm data}} )^2}
                                { \sigma_{b,\,{\rm sys}}^2} } \,.
\ee
The probability-to-exceed (PTE) of this $\chi^2$ is calculated from a $\chi^2$ distribution with \nbin{} degrees of freedom (corresponding to the \nbin{} bins in our spectrum).

The second metric considers the change in the lensing amplitude.
We calculate the lensing difference-amplitudes ($\Delta A_{\rm \mv}$) as defined in Equation \ref{eqn:sys} for the data and for each simulation.
We then calculate the variance of the simulation difference-amplitudes (${\rm var}(\Delta A_{\rm MV, sim})$);
this variance estimates the expected magnitude of the change in the lensing amplitude.
Specifically, we expect the magnitude of the lensing difference-amplitude in the data to be less than or equal to $\sqrt{{\rm var}(\Delta A_{\rm MV, sim})}$ in $\sim 68\%$ of similar measurements.
Finally, we calculate the PTE of the data difference-amplitude directly from the simulations as the percentage of simulations that have a difference-amplitude with a larger magnitude than $\Delta A_{\rm \mv}$ for the data.
The data difference-amplitude, variance of simulated difference-amplitudes, and PTEs for each test are shown in Table \ref{tab:systematics}.

The individual tests are described in detail below, 
and the results are reported in Table \ref{tab:systematics} and Figure \ref{fig:systematics}.

\begin{enumerate}

  \item{$\pmb{\ell_{xmin}}$ \textbf{cut}: As described in Section \ref{sec:data}, we cut all modes with $|\ell_x| < 450$ from the CMB maps.  
    In this test, we adjust that cut from $|\ell_x| < 450$ to 400 and 500.
    When more $\ell$-space is removed by increasing the $|\ell_x|$ cut to 500, the change in the lensing spectrum and amplitude are consistent with expectations from simulations.
    On the other hand, including the region between $400 < |\ell_x| < 450$ causes an unexpectedly large shift ($>2\sigma$) in the lensing amplitude, thus motivating the placement of this cut.
    Even in this case, however, the change in the amplitude is only $0.4\sigma$ with respect to the statistical uncertainty on the lensing spectrum amplitude.
}

  \item{$\pmb{\ell_{max}}$ \textbf{cut}: We adjust the maximum value of $\ell$ from CMB maps that is used in the estimator from $\ell<3000$ to 2500 and 3500.
    This corresponds to adjusting the upper bound of the integral in Equation \ref{eq:phi_bar}.
    We find the data are consistent with the expectation from simulations in this test.}

  \item{\textbf{Apodized Mask}: We apodize the sky and point-source mask with a cosine profile on the edges, and recalculate the amplitude and \clpp{}.
    The change in amplitude is consistent with the expectation from simulations.}

  \item{\textbf{C-inverse test}: We recalculate the \mv{} spectrum and amplitude using the diagonal approximation to the C-inverse filter 
    (e.g., $\mathcal{F}^{X}_{\mbell}$ from Equation \ref{eq:response}).
    The covariance of the simulations increases by a factor of 
    four
    over the nominal covariance (due in part to the fact that the C-inverse filter effectively apodizes the map, while the diagonal approximation is just an inverse-variance filter with no apodization),
    but the shift in the lensing amplitude is consistent with the shift of the simulations.}

  \item{\textbf{Curl test}: We replace the gradient estimator with a curl estimator that is optimized for curl-like sources \citep{cooray05}.
      Specifically, we use the curl estimators specified in Table 3 of \cite{namikawa12}.
      We estimate and remove the N0 contribution to the curl estimator as in Equation \ref{eq:n1}; 
      thus the curl test provides a second check of the N0 estimation procedure.
      A non-zero curl signal would indicate contamination in the nominal gradient reconstruction from non-Gaussian secondary effects or foregrounds \citep{cooray05}.
      We calculate the curl signal, and find it to be consistent with the expectation from simulations in both the 
      \mv{} estimator ($\Delta A_{\rm \mv} = 0.082 \pm 0.092$, \clpp{} Spectrum PTE = 0.69) and the 
      \pp{} estimator ($\Delta A_{\rm \pp} = 0.055 \pm 0.110$, \clpp{} Spectrum PTE = 0.52).
  }

  \item{\textbf{Scan Direction}: We perform a ``jackknife'' null test on the telescope scan direction.
    This test is sensitive to any systematic differences between left-going vs. right-going scans. 
    We calculate null maps by subtracting all left-going scans from all right-going scans.
    The resulting maps should be free from signal but still contain any systematic difference between left-going and right-going maps.
    Because this is a null test, we calculate the pass-fail metrics for the jackknife spectrum and amplitude relative to zero (rather than relative to the \mv{} spectrum and amplitude).
    Formally, this means replacing the baseline spectrum $C^{\phi \phi}_b$ and amplitude $A_{\rm \mv}$ in Equation \ref{eqn:sys} with zeros, and using noise-only simulations.
    We find that the null spectrum and amplitude are consistent with noise.
}

\end{enumerate}

Finally, we compare the spectra and amplitudes from each of the estimators with those from the MV spectrum.
We find that the spectra are mostly consistent.
The ``TExEB'' spectrum amplitude is high by $\sim 2.1\sigma$ relative to the expectation from simulations, 
while the ``EExEE'' spectrum amplitude is low by $\sim 2.8\sigma$.
All other spectra are consistent to within $2\sigma$.

\section{Discussion}
\label{sec:discussion}

This paper presents a measurement of the CMB lensing potential $\phi$ from 100 deg$^2$ of sky observed in temperature and polarization with \sptpol.
Using a quadratic estimator analysis including polarization information, we have constructed a map of the lensing convergence field.
Individual Fourier modes in this map are measured with signal-to-noise greater than one in the angular wavenumber or multipole range $100 < L < 250.$
This represents the highest signal-to-noise map of the integrated lensing potential made from the CMB to date.
The power spectrum of the lensing potential \clpp{} was calculated from these maps.
We have verified that this measurement is robust against systematics by performing a suite of systematics and null tests.

We compare this measurement to a fiducial spectrum taken from the \textsc{Planck+Lens+WP+highL} best-fit \LCDM{} model and find a relative amplitude of 
$A_{\rm \mv}=\mvAmp \pm \mvAmpStat {\rm\, (Stat.)} \pm \mvAmpSys {\rm\, (Sys.)}$.
This corresponds to a \mvTotPercent{} measurement of the amplitude.
This measurement rejects the no-lensing hypothesis at \mvSigUnl{}.
If instead only polarized estimators are used, we find
$A_{\rm \pp}=\ppAmp \pm \ppAmpStat {\rm\, (Stat.)} \pm \ppAmpSys {\rm\, (Sys.)}$.
This is a \ppTotPercent{} measurement of the amplitude.
This measurement rejects the no-lensing hypothesis using polarization information only at \ppSigUnl{}.
The \mv{} and \pp{} amplitudes are consistent at \mvLCDMSig{} and \ppLCDMSig{} , respectively, with the best-fit \LCDM{} cosmology of the \textsc{Planck+Lens+WP+highL} dataset.
The PTE of the $\chi^2$ relative to this theory spectrum for the \mv{} and \pp{} spectra are \mvPTE{} and \ppPTE{}, respectively.
The PTEs for all individual estimators are nominal, with a max/min values of 0.96 and 0.16, respectively.
Of note, the bandpowers are consistent with the \LCDM{} theory prediction even at high $L$, where non-linear structure growth or contamination from foregrounds are more relevant.

We also compare the amplitude of our measured lensing spectrum relative to other cosmologies.
Within the context of a given cosmological model, the true amplitude of the lensing spectrum is by definition unity; 
any statistically significant deviation of the measured value of $A_{\rm lens}$ from unity indicates tension with the assumed cosmological model.
Replacing our fiducial lensing spectrum with the lensing spectrum derived from the \LCDM{} model that best fits the \textsc{WMAP9+SPT-SZ} dataset 
(a combination of \textit{WMAP}9 data \citep{hinshaw13} and temperature power spectrum from \citealt{story13}), 
the best-fit amplitude is
$A_{\rm \mv}^{\rm WMAP9+SPT} = 1.05$.
This represents a shift of $\sim 13\%$, or $\sim 1\sigma$ (statistical-only) from the best-fit value assuming our fiducial cosmology. 
The current \sptpol{} measurement is consistent with the predictions of both cosmologies and cannot distinguish between them.

In comparison with previous quadratic estimator measurements from SPT and ACT, which used roughly six times the sky area ($\sim 600$ deg$^2$), we have measured the lensing potential with similar precision\footnote{
The fractional precision is reported relative to the mean of each measurement.}: 
\mvTotPercent{} in this measurement, as compared to 
$19\%$ from \cite{vanengelen12} and 
$22\%$ from \citealt{das14}. 
The precision of our measurement, however, is limited by sample-variance of the lenses themselves;
as a result, our measurement rules out the no-lensing hypothesis much more significantly, at 
\mvSigUnl{} in our \mv{} measurement.
For comparison, no lensing was ruled out at $6.3\sigma$ in \cite{vanengelen12} (although the sample variance was scaled as a function of $A_L$ in that calculation) and at $>4.6\sigma$ in \cite{das14} (where sample variance was included).
The temperature measurement from \planck{} is derived from over $70\%$ of the sky and thus has much lower sample variance. 
At $25\sigma$ it is highly significant, although no modes are measured with a signal-to-noise ratio greater than one \citep{planck13-17}.
Our measurement is consistent with that from \planck{};
we calculate the \mv{} amplitude relative to the same fiducial cosmology used in \cite{planck13-17}, and find an amplitude of $A_{\rm \mv}^{\rm \planck-Fiducial} = 0.90 \pm \mvDampTot$, which is consistent with the measurement from \planck{} of $A_{\rm \planck}=0.94 \pm 0.04$.
Finally, our polarization-only measurement improves on the sensitivity achieved by \polarbear{}: 
the \sptpol{} \pp{} spectrum rules out no-lensing at \ppSigUnl{}, as compared to $4.2\sigma$ from (\citeauthor{polarbear2014a}).

Because of the high signal-to-noise of the \sptpol{} measurement presented here
and the fact that the \bicep{} and \keck{} experiments have observed the same patch of sky,
this \sptpol{} mass map will be powerful for de-lensing primordial BB power spectrum measurements.
Additionally, these deep sptpol mass maps will enable significant cross-correlation measurements with other tracers of large-scale structure.

Polarization measurements will continue to improve rapidly.
\sptpol{} will soon start its fourth year of observing a larger 500 deg$^2$ patch of sky,
ACTpol and \polarbear{} observations are continuing,
and \planck{} is expected to release a similar polarized analysis in the near future. 
With these and the next generation of CMB polarization experiments being planned
(e.g., SPT3G \citep{benson14}, AdvancedACTpol \citep{calabrese14}, Simons Array \citep{arnold14}), 
CMB lensing will become an exceptionally powerful probe of structure evolution in the universe.

\acknowledgements{
The South Pole Telescope program is supported by the National Science Foundation through grant PLR-1248097. Partial support is also provided by the NSF Physics Frontier Center grant PHY-0114422 to the Kavli Institute of Cosmological Physics at the University of Chicago, the Kavli Foundation, and the Gordon and Betty Moore Foundation through Grant GBMF\#947 to the University of Chicago.  
The McGill authors acknowledge funding from the Natural Sciences and Engineering Research Council of Canada, Canadian Institute for Advanced Research, and Canada Research Chairs program.
The CU Boulder group acknowledges support from NSF AST-0956135.  
JWH is supported by the National Science Foundation under Award No. AST-1402161.
BB is supported by the Fermi Research Alliance, LLC under Contract No. De-AC02-07CH11359 with the U.S. Department of Energy.  
TdH is supported by a Miller Research Fellowship.
This work is also supported by the U.S. Department of Energy.  
Work at Argonne National Lab is supported by UChicago Argonne, LLC, Operator of Argonne National Laboratory (Argonne). 
Argonne, a U.S. Department of Energy Office of Science Laboratory, is operated under Contract No. DE-AC02-06CH11357. 
We also acknowledge support from the Argonne Center for Nanoscale Materials.  
This research used resources of the Calcul Quebec computing consortium, part of the Compute Canada network,
and of the National Energy Research Scientific Computing Center, a DOE Office of Science User Facility supported by the Office of Science of the U.S. Department of Energy under Contract No. DE-AC02-05CH11231.
The data analysis pipeline uses the scientific Python stack \citep{hunter07, jones01, vanDerWalt11} and the HDF5 file format \citep{hdf5}.
}

\bibliography{../../BIBTEX/spt}

\end{document}

%% file: lens100d_authorlist_v1.tex
\def\KICPChicago{1}
\def\PhysicsUChicago{2}
\def\McGill{3}
\def\Cardiff{4}
\def\UChicago{5}
\def\ColoradoAPS{6}
\def\NIST{7}
\def\ArgonneHEP{8}
\def\AAUChicago{9}
\def\FNAL{10}
\def\EFIChicago{11}
\def\UKZN{12}
\def\SLAC{13}
\def\Caltech{14}
\def\Berkeley{15}
\def\CIFAR{16}
\def\ColoradoPhys{17}
\def\Stanford{18}
\def\KIPAC{19}
\def\Davis{20}
\def\LBNL{21}
\def\Michigan{22}
\def\CaseWestern{23}
\def\ArgonneMSD{24}
\def\Minnesota{25}
\def\Melbourne{26}
\def\ArtInstChicago{27}
\def\ThreeSpeedLogic{28}
\def\CfA{29}
\def\Dunlap{30}
\def\UToronto{31}
\def\illast{32}
\def\illphy{33}
\def\BCCP{34}


 \author{
  K.~T.~Story\altaffilmark{\KICPChicago,\PhysicsUChicago},
  D.~Hanson\altaffilmark{\McGill},
  P.~A.~R.~Ade\altaffilmark{\Cardiff},
  K.~A.~Aird\altaffilmark{\UChicago},
  J.~E.~Austermann\altaffilmark{\ColoradoAPS},
  J.~A.~Beall\altaffilmark{\NIST} ,
  A.~N.~Bender\altaffilmark{\McGill,\ArgonneHEP},
  B.~A.~Benson\altaffilmark{\KICPChicago,\AAUChicago,\FNAL},
  L.~E.~Bleem\altaffilmark{\KICPChicago,\PhysicsUChicago,\ArgonneHEP},
  J.~E.~Carlstrom\altaffilmark{\KICPChicago,\PhysicsUChicago,\ArgonneHEP,\AAUChicago,\EFIChicago},
  C.~L.~Chang\altaffilmark{\KICPChicago,\ArgonneHEP,\AAUChicago},
  H.~C.~Chiang\altaffilmark{\UKZN},
  H-M.~Cho\altaffilmark{\SLAC},
  R.~Citron\altaffilmark{\KICPChicago},
  T.~M.~Crawford\altaffilmark{\KICPChicago,\AAUChicago},
  A.~T.~Crites\altaffilmark{\KICPChicago,\AAUChicago,\Caltech},
  T.~de~Haan\altaffilmark{\Berkeley}, 
  M.~A.~Dobbs\altaffilmark{\McGill,\CIFAR},
  W.~Everett\altaffilmark{\ColoradoAPS},
  J.~Gallicchio\altaffilmark{\KICPChicago},
  J.~Gao\altaffilmark{\NIST},
  E.~M.~George\altaffilmark{\Berkeley},
  A.~Gilbert\altaffilmark{\McGill},
  N.~W.~Halverson\altaffilmark{\ColoradoAPS,\ColoradoPhys},
  N.~Harrington\altaffilmark{\Berkeley},
  J.~W.~Henning\altaffilmark{\KICPChicago,\ColoradoAPS},
  G.~C.~Hilton\altaffilmark{\NIST},
  G.~P.~Holder\altaffilmark{\McGill},
  W.~L.~Holzapfel\altaffilmark{\Berkeley},
  S.~Hoover\altaffilmark{\KICPChicago,\PhysicsUChicago},
  Z.~Hou\altaffilmark{\KICPChicago},
  J.~D.~Hrubes\altaffilmark{\UChicago},
  N.~Huang\altaffilmark{\Berkeley},
  J.~Hubmayr\altaffilmark{\NIST},
  K.~D.~Irwin\altaffilmark{\SLAC,\Stanford},
  R.~Keisler\altaffilmark{\Stanford,\KIPAC},
  L.~Knox\altaffilmark{\Davis},
  A.~T.~Lee\altaffilmark{\Berkeley,\LBNL},
  E.~M.~Leitch\altaffilmark{\KICPChicago,\AAUChicago},
  D.~Li\altaffilmark{\NIST,\SLAC},
  C.~Liang\altaffilmark{\UChicago},       
  D.~Luong-Van\altaffilmark{\UChicago},
  J.~J.~McMahon\altaffilmark{\Michigan},
  J.~Mehl\altaffilmark{\KICPChicago,\ArgonneHEP},
  S.~S.~Meyer\altaffilmark{\KICPChicago,\PhysicsUChicago,\AAUChicago,\EFIChicago},
  L.~Mocanu\altaffilmark{\KICPChicago,\AAUChicago},
  T.~E.~Montroy\altaffilmark{\CaseWestern},
  T.~Natoli\altaffilmark{\KICPChicago,\PhysicsUChicago},
  J.~P.~Nibarger\altaffilmark{\NIST},
  V.~Novosad\altaffilmark{\ArgonneMSD},
  S.~Padin\altaffilmark{\KICPChicago,\AAUChicago,\Caltech},
  C.~Pryke\altaffilmark{\Minnesota},
  C.~L.~Reichardt\altaffilmark{\Berkeley,\Melbourne},
  J.~E.~Ruhl\altaffilmark{\CaseWestern},
  B.~R.~Saliwanchik\altaffilmark{\CaseWestern},
  J.T.~Sayre\altaffilmark{\CaseWestern},
  K.~K.~Schaffer\altaffilmark{\KICPChicago,\EFIChicago,\ArtInstChicago},
  G.~Smecher\altaffilmark{\McGill,\ThreeSpeedLogic},
  A.~A.~Stark\altaffilmark{\CfA},
  C.~Tucker\altaffilmark{\Cardiff},
  K.~Vanderlinde\altaffilmark{\Dunlap,\UToronto},
  J.~D.~Vieira\altaffilmark{\illast,\illphy},
  G.~Wang\altaffilmark{\ArgonneHEP},
  N.~Whitehorn\altaffilmark{\Berkeley},
  V.~Yefremenko\altaffilmark{\ArgonneHEP},
  and 
  O.~Zahn\altaffilmark{\BCCP}
 }

\altaffiltext{\KICPChicago}{Kavli Institute for Cosmological Physics, University of Chicago, 5640 South Ellis Avenue, Chicago, IL, USA 60637}
\altaffiltext{\PhysicsUChicago}{Department of Physics, University of Chicago, 5640 South Ellis Avenue, Chicago, IL, USA 60637}
\altaffiltext{\McGill}{Department of Physics, McGill University, 3600 Rue University, Montreal, Quebec H3A 2T8, Canada}
\altaffiltext{\Cardiff}{Cardiff University, Cardiff CF10 3XQ, United Kingdom}
\altaffiltext{\UChicago}{University of Chicago, 5640 South Ellis Avenue, Chicago, IL, USA 60637}
\altaffiltext{\ColoradoAPS}{Department of Astrophysical and Planetary Sciences, University of Colorado, Boulder, CO, USA 80309}
\altaffiltext{\NIST}{NIST Quantum Devices Group, 325 Broadway Mailcode 817.03, Boulder, CO, USA 80305}
\altaffiltext{\ArgonneHEP}{High Energy Physics Division, Argonne National Laboratory,9700 S. Cass Avenue, Argonne, IL, USA 60439}
\altaffiltext{\AAUChicago}{Department of Astronomy and Astrophysics, University of Chicago, 5640 South Ellis Avenue, Chicago, IL, USA 60637}
\altaffiltext{\FNAL}{Fermi National Accelerator Laboratory, MS209, P.O. Box 500, Batavia, IL 60510}
\altaffiltext{\EFIChicago}{Enrico Fermi Institute, University of Chicago, 5640 South Ellis Avenue, Chicago, IL, USA 60637}
\altaffiltext{\UKZN}{School of Mathematics, Statistics \& Computer Science, University of KwaZulu-Natal, Durban, South Africa}
\altaffiltext{\SLAC}{SLAC National Accelerator Laboratory, 2575 Sand Hill Road, Menlo Park, CA 94025}
\altaffiltext{\Caltech}{California Institute of Technology, MS 249-17, 1216 E. California Blvd., Pasadena, CA, USA 91125}
\altaffiltext{\Berkeley}{Department of Physics, University of California, Berkeley, CA, USA 94720}
\altaffiltext{\CIFAR}{Canadian Institute for Advanced Research, CIFAR Program in Cosmology and Gravity, Toronto, ON, M5G 1Z8, Canada}
\altaffiltext{\ColoradoPhys}{Department of Physics, University of Colorado, Boulder, CO, USA 80309}
\altaffiltext{\Stanford}{Dept. of Physics, Stanford University, 382 Via Pueblo Mall, Stanford, CA 94305}
\altaffiltext{\KIPAC}{Kavli Institute for Particle Astrophysics and Cosmology, Stanford University, 452 Lomita Mall, Stanford, CA 94305}
\altaffiltext{\Davis}{Department of Physics, University of California, One Shields Avenue, Davis, CA, USA 95616}
\altaffiltext{\LBNL}{Physics Division, Lawrence Berkeley National Laboratory, Berkeley, CA, USA 94720}
\altaffiltext{\Michigan}{Department of Physics, University of Michigan, 450 Church Street, Ann  Arbor, MI, USA 48109}
\altaffiltext{\CaseWestern}{Physics Department, Center for Education and Research in Cosmology and Astrophysics, Case Western Reserve University,Cleveland, OH, USA 44106}
\altaffiltext{\ArgonneMSD}{Materials Sciences Division, Argonne National Laboratory,9700 S. Cass Avenue, Argonne, IL, USA 60439}
\altaffiltext{\Minnesota}{School of Physics and Astronomy, University of Minnesota, 116 Church Street S.E. Minneapolis, MN, USA 55455}
\altaffiltext{\Melbourne}{School of Physics, University of Melbourne, Parkville, VIC 3010, Australia}
\altaffiltext{\ArtInstChicago}{Liberal Arts Department, School of the Art Institute of Chicago, 112 S Michigan Ave, Chicago, IL, USA 60603}
\altaffiltext{\ThreeSpeedLogic}{Three-Speed Logic, Inc., Vancouver, B.C., V6A 2J8, Canada}
\altaffiltext{\CfA}{Harvard-Smithsonian Center for Astrophysics, 60 Garden Street, Cambridge, MA, USA 02138}
\altaffiltext{\Dunlap}{Dunlap Institute for Astronomy \& Astrophysics, University of Toronto, 50 St George St, Toronto, ON, M5S 3H4, Canada}
\altaffiltext{\UToronto}{Department of Astronomy \& Astrophysics, University of Toronto, 50 St George St, Toronto, ON, M5S 3H4, Canada}
\altaffiltext{\illast}{Astronomy Department, University of Illinois at Urbana-Champaign, 1002 W.\ Green Street, Urbana, IL 61801, USA}
\altaffiltext{\illphy}{Department of Physics, University of Illinois Urbana-Champaign, 1110 W.\ Green Street, Urbana, IL 61801, USA}
\altaffiltext{\BCCP}{Berkeley Center for Cosmological Physics, Department of Physics, University of California, and Lawrence Berkeley National Laboratory, Berkeley, CA, USA 94720}

\email{kstory@uchicago.edu}